\documentclass[aps,prd,showpacs,superscriptaddress,twocolumn,floatfix]{revtex4} 

\usepackage{graphicx}
\usepackage{natbib}
\usepackage{slashed}
\usepackage{graphics}
\usepackage{graphicx}
\usepackage{setspace}
\usepackage[scriptsize,tight,nooneline]{subfigure}
\usepackage{amssymb}
\usepackage{amsmath}
\usepackage{fancyhdr}
\usepackage{color}
\usepackage{rotating}
\usepackage{lineno}
\usepackage{multirow}
\usepackage{setspace}
\usepackage{relsize}
\usepackage{lscape}

\newcommand{\dzSVT}{\ensuremath{d_0^{\mathrm{L2}}}}
\newcommand{\dzLt}{\ensuremath{d_0^{\mathrm{L3}}}}
\newcommand{\dzOff}{\ensuremath{d_0^{\mathrm{off}}}}
\newcommand{\delDz}{\ensuremath{\Delta d_0}}

\newcommand{\effSVT}{\ensuremath{\varepsilon_s}}
\newcommand{\eff}{\ensuremath{\varepsilon}}
\newcommand{\effbkg}{\ensuremath{{\varepsilon_s^{\mathrm{bkg}}}}}
\newcommand{\effbg}{\effbkg}
\newcommand{\effb}{\effbkg}
\newcommand{\effsg}{\ensuremath{\varepsilon_s^{\mathrm{sig}}}}

\newcommand{\tm}{\ensuremath{t_i} }

\newcommand{\mass}{\ensuremath{m_i}}
\newcommand{\acct}{\ensuremath{E_i(t)}}
\newcommand{\acc}{\ensuremath{E_i(t, \effSVT)}}
\newcommand{\accb}{\ensuremath{E_i(t, \effb)}}

\newcommand{\trk}{\ensuremath{C_i}}
\newcommand{\trig}{\ensuremath{\mathrm{T}}}

\newcommand{\intervSum}{\ensuremath{
\sum\limits_{\parbox{3.5em}{\scriptsize\center \mbox{}\vspace{-4ex}\\$k_i$=all\vspace{-0.5ex}\newline intervals\vspace{-0.5ex}\newline in event $i$}}}}

\newcommand{\tmin}{\ensuremath{t_{\mathrm{min}}}}
\newcommand{\tmax}{\ensuremath{t_{\mathrm{max}}}}

\newcommand{\gauss}[2]{\ensuremath{ \frac{1}{\sqrt{2\pi}#2} e^{-\frac{{#1}^2}{2{#2}^2}}   }}

\newcommand{\Freq}{\ensuremath{{F}}}

\newcommand{\half}{\ensuremath{\frac{1}{2}}}

\newcommand{\un}[2]{\ensuremath{#1\,\mathrm{#2}}}

\def\bplus {$B^{-}$ }

\def\d0{$D^0$ }

\newcommand{\btodpinogap}{$B^{-}\rightarrow D^0 \pi^-$}
\newcommand{\btodpi}{\btodpinogap{} }
\newcommand{\bdknogap}{$B^{-} \rightarrow D^0 K^{-}$}
\newcommand{\bdk}{\bdknogap{} }
\newcommand{\dtopiknogap}{$D^0 \rightarrow K^{-} \pi^{+}$}
\newcommand{\dtopik}{\dtopik{} }
\def\mum{$\mu$m }
\begin{document}

\newcommand{\rotAngle}{270}
\newcommand{\vect}[1]{\boldsymbol{#1}}

\normalsize
\textwidth=15.5cm
\newcommand{\pw}{height}

\newcommand{\secref}[1]{Sec.~\ref{#1}}
\newcommand{\Secref}[1]{Sec.~\ref{#1}}
\newcommand{\appref}[1]{Appendix~\ref{#1}}
\newcommand{\Appref}[1]{Appendix~\ref{#1}}
\newcommand{\figref}[1]{Fig.~\ref{#1}}
\newcommand{\Figref}[1]{Figure~\ref{#1}}
\newcommand{\tabref}[1]{Table~\ref{#1}}
\newcommand{\Tabref}[1]{Table~\ref{#1}}
\newcommand{\equationref}[1]{Eq.~(\ref{#1})}
\newcommand{\Equationref}[1]{Equation~(\ref{#1})}

\newcommand{\geantIII}{{\sc geant~3}}
\newcommand{\geant}{{\sc geant}}

\pagestyle{plain}

\bibliographystyle{apsrev}

%%%%%\preprint{CDF/DOC/TOP/CDFR/WXYZ} %Comment this out before shipping out
%\preprint{Phys. Rev. D draft 3.0}
%\vspace*{-1.5cm}
%\includegraphics[width=0.15\textwidth]{Pictures/cdf_ii.eps}
%\vspace{1.5cm}

\title{Measurement of the $B^-$ lifetime using a simulation free approach for trigger bias correction}

\vspace{0.5in}

\affiliation{Institute of Physics, Academia Sinica, Taipei, Taiwan 11529, Republic of China} 
\affiliation{Argonne National Laboratory, Argonne, Illinois 60439, USA} 
\affiliation{University of Athens, 157 71 Athens, Greece} 
\affiliation{Institut de Fisica d'Altes Energies, Universitat Autonoma de Barcelona, E-08193, Bellaterra (Barcelona), Spain} 
\affiliation{Baylor University, Waco, Texas 76798, USA} 
\affiliation{Istituto Nazionale di Fisica Nucleare Bologna, $^{dd}$University of Bologna, I-40127 Bologna, Italy} 
\affiliation{Brandeis University, Waltham, Massachusetts 02254, USA} 
\affiliation{University of California, Davis, Davis, California 95616, USA} 
\affiliation{University of California, Los Angeles, Los Angeles, California 90024, USA} 
\affiliation{University of California, San Diego, La Jolla, California 92093, USA} 
\affiliation{University of California, Santa Barbara, Santa Barbara, California 93106, USA} 
\affiliation{Instituto de Fisica de Cantabria, CSIC-University of Cantabria, 39005 Santander, Spain} 
\affiliation{Carnegie Mellon University, Pittsburgh, Pennsylvania 15213, USA} 
\affiliation{Enrico Fermi Institute, University of Chicago, Chicago, Illinois 60637, USA}
\affiliation{Comenius University, 842 48 Bratislava, Slovakia; Institute of Experimental Physics, 040 01 Kosice, Slovakia} 
\affiliation{Joint Institute for Nuclear Research, RU-141980 Dubna, Russia} 
\affiliation{Duke University, Durham, North Carolina 27708, USA} 
\affiliation{Fermi National Accelerator Laboratory, Batavia, Illinois 60510, USA} 
\affiliation{University of Florida, Gainesville, Florida 32611, USA} 
\affiliation{Laboratori Nazionali di Frascati, Istituto Nazionale di Fisica Nucleare, I-00044 Frascati, Italy} 
\affiliation{University of Geneva, CH-1211 Geneva 4, Switzerland} 
\affiliation{Glasgow University, Glasgow G12 8QQ, United Kingdom} 
\affiliation{Harvard University, Cambridge, Massachusetts 02138, USA} 
\affiliation{Division of High Energy Physics, Department of Physics, University of Helsinki and Helsinki Institute of Physics, FIN-00014, Helsinki, Finland} 
\affiliation{University of Illinois, Urbana, Illinois 61801, USA} 
\affiliation{The Johns Hopkins University, Baltimore, Maryland 21218, USA} 
\affiliation{Institut f\"{u}r Experimentelle Kernphysik, Karlsruhe Institute of Technology, D-76131 Karlsruhe, Germany} 
\affiliation{Center for High Energy Physics: Kyungpook National University, Daegu 702-701, Korea; Seoul National University, Seoul 151-742, Korea; Sungkyunkwan University, Suwon 440-746, Korea; Korea Institute of Science and Technology Information, Daejeon 305-806, Korea; Chonnam National University, Gwangju 500-757, Korea; Chonbuk National University, Jeonju 561-756, Korea} 
\affiliation{Ernest Orlando Lawrence Berkeley National Laboratory, Berkeley, California 94720, USA} 
\affiliation{University of Liverpool, Liverpool L69 7ZE, United Kingdom} 
\affiliation{University College London, London WC1E 6BT, United Kingdom} 
\affiliation{Centro de Investigaciones Energeticas Medioambientales y Tecnologicas, E-28040 Madrid, Spain} 
\affiliation{Massachusetts Institute of Technology, Cambridge, Massachusetts 02139, USA} 
\affiliation{Institute of Particle Physics: McGill University, Montr\'{e}al, Qu\'{e}bec, Canada H3A~2T8; Simon Fraser University, Burnaby, British Columbia, Canada V5A~1S6; University of Toronto, Toronto, Ontario, Canada M5S~1A7; and TRIUMF, Vancouver, British Columbia, Canada V6T~2A3} 
\affiliation{University of Michigan, Ann Arbor, Michigan 48109, USA} 
\affiliation{Michigan State University, East Lansing, Michigan 48824, USA}
\affiliation{Institution for Theoretical and Experimental Physics, ITEP, Moscow 117259, Russia} 
\affiliation{University of New Mexico, Albuquerque, New Mexico 87131, USA} 
\affiliation{Northwestern University, Evanston, Illinois 60208, USA} 
\affiliation{The Ohio State University, Columbus, Ohio 43210, USA} 
\affiliation{Okayama University, Okayama 700-8530, Japan} 
\affiliation{Osaka City University, Osaka 588, Japan} 
\affiliation{University of Oxford, Oxford OX1 3RH, United Kingdom} 
\affiliation{Istituto Nazionale di Fisica Nucleare, Sezione di Padova-Trento, $^{ee}$University of Padova, I-35131 Padova, Italy} 
\affiliation{LPNHE, Universite Pierre et Marie Curie/IN2P3-CNRS, UMR7585, Paris, F-75252 France} 
\affiliation{University of Pennsylvania, Philadelphia, Pennsylvania 19104, USA}
\affiliation{Istituto Nazionale di Fisica Nucleare Pisa, $^{ff}$University of Pisa, $^{gg}$University of Siena and $^{hh}$Scuola Normale Superiore, I-56127 Pisa, Italy} 
\affiliation{University of Pittsburgh, Pittsburgh, Pennsylvania 15260, USA} 
\affiliation{Purdue University, West Lafayette, Indiana 47907, USA} 
\affiliation{University of Rochester, Rochester, New York 14627, USA} 
\affiliation{The Rockefeller University, New York, New York 10021, USA} 
\affiliation{Istituto Nazionale di Fisica Nucleare, Sezione di Roma 1, $^{ii}$Sapienza Universit\`{a} di Roma, I-00185 Roma, Italy} 

\affiliation{Rutgers University, Piscataway, New Jersey 08855, USA} 
\affiliation{Texas A\&M University, College Station, Texas 77843, USA} 
\affiliation{Istituto Nazionale di Fisica Nucleare Trieste/Udine, I-34100 Trieste, $^{jj}$University of Trieste/Udine, I-33100 Udine, Italy} 
\affiliation{University of Tsukuba, Tsukuba, Ibaraki 305, Japan} 
\affiliation{Tufts University, Medford, Massachusetts 02155, USA} 
\affiliation{Waseda University, Tokyo 169, Japan} 
\affiliation{Wayne State University, Detroit, Michigan 48201, USA} 
\affiliation{University of Wisconsin, Madison, Wisconsin 53706, USA} 
\affiliation{Yale University, New Haven, Connecticut 06520, USA} 
\author{T.~Aaltonen}
\affiliation{Division of High Energy Physics, Department of Physics, University of Helsinki and Helsinki Institute of Physics, FIN-00014, Helsinki, Finland}
\author{J.~Adelman}
\affiliation{Enrico Fermi Institute, University of Chicago, Chicago, Illinois 60637, USA}
\author{B.~\'{A}lvarez~Gonz\'{a}lez$^w$}
\affiliation{Instituto de Fisica de Cantabria, CSIC-University of Cantabria, 39005 Santander, Spain}
\author{S.~Amerio$^{ee}$}
\affiliation{Istituto Nazionale di Fisica Nucleare, Sezione di Padova-Trento, $^{ee}$University of Padova, I-35131 Padova, Italy} 

\author{D.~Amidei}
\affiliation{University of Michigan, Ann Arbor, Michigan 48109, USA}
\author{A.~Anastassov}
\affiliation{Northwestern University, Evanston, Illinois 60208, USA}
\author{A.~Annovi}
\affiliation{Laboratori Nazionali di Frascati, Istituto Nazionale di Fisica Nucleare, I-00044 Frascati, Italy}
\author{J.~Antos}
\affiliation{Comenius University, 842 48 Bratislava, Slovakia; Institute of Experimental Physics, 040 01 Kosice, Slovakia}
\author{G.~Apollinari}
\affiliation{Fermi National Accelerator Laboratory, Batavia, Illinois 60510, USA}
\author{J.~Appel}
\affiliation{Fermi National Accelerator Laboratory, Batavia, Illinois 60510, USA}
% ...

\collaboration{CDF Collaboration\footnote{With visitors from $^a$University of Massachusetts Amherst, Amherst, Massachusetts 01003,
$^b$Universiteit Antwerpen, B-2610 Antwerp, Belgium, 
$^c$University of Bristol, Bristol BS8 1TL, United Kingdom,
$^d$Chinese Academy of Sciences, Beijing 100864, China, 
$^e$Istituto Nazionale di Fisica Nucleare, Sezione di Cagliari, 09042 Monserrato (Cagliari), Italy,
$^f$University of California Irvine, Irvine, CA  92697, 
$^g$University of California Santa Cruz, Santa Cruz, CA  95064, 
$^h$Cornell University, Ithaca, NY  14853, 
$^i$University of Cyprus, Nicosia CY-1678, Cyprus, 
$^j$University College Dublin, Dublin 4, Ireland,
$^k$University of Edinburgh, Edinburgh EH9 3JZ, United Kingdom, 
$^l$University of Fukui, Fukui City, Fukui Prefecture, Japan 910-0017,
$^m$Kinki University, Higashi-Osaka City, Japan 577-8502,
$^n$Universidad Iberoamericana, Mexico D.F., Mexico,
$^o$University of Iowa, Iowa City, IA  52242,
$^p$Kansas State University, Manhattan, KS 66506,
$^q$Queen Mary, University of London, London, E1 4NS, England,
$^r$University of Manchester, Manchester M13 9PL, England,
$^s$Muons, Inc., Batavia, IL 60510, 
$^t$Nagasaki Institute of Applied Science, Nagasaki, Japan, 
$^u$University of Notre Dame, Notre Dame, IN 46556,
$^v$Obninsk State University, Obninsk, Russia,
$^w$University de Oviedo, E-33007 Oviedo, Spain, 
$^x$Texas Tech University, Lubbock, TX  79609, 
$^y$IFIC(CSIC-Universitat de Valencia), 56071 Valencia, Spain,
$^z$Universidad Tecnica Federico Santa Maria, 110v Valparaiso, Chile,
$^{aa}$University of Virginia, Charlottesville, VA  22906,
$^{bb}$Bergische Universit\"at Wuppertal, 42097 Wuppertal, Germany,
$^{cc}$Yarmouk University, Irbid 211-63, Jordan,
$^{kk}$On leave from J.~Stefan Institute, Ljubljana, Slovenia, 
}}
\noaffiliation

\date{\today}

\begin{abstract}
The collection of a large number of $B$ hadron decays to hadronic
final states at the CDF II detector is possible due to the presence of
a trigger that selects events based on track impact
parameters. However, the nature of the selection requirements of the
trigger introduces a large bias in the observed proper decay time
distribution. A lifetime measurement must correct for this bias and
the conventional approach has been to use a Monte Carlo
simulation. The leading sources of systematic uncertainty in the
conventional approach are due to differences between the data and the
Monte Carlo simulation. In this paper we present an analytic method
for bias correction without using simulation, thereby removing any
uncertainty due to the differences between data and simulation.  This method is presented in
the form of a measurement of the lifetime of the \bplus using the mode
\btodpinogap. The $B^-$ lifetime is measured as $\tau_{B^-}$ = 1.663
$\pm$ 0.023 $\pm$ 0.015 ps, where the first uncertainty is statistical
and the second systematic. This new method results in a smaller
systematic uncertainty in comparison to methods that use simulation to
correct for the trigger bias.

\end{abstract}

% 

\maketitle


%------------------------------------------------------------

\section{Introduction}
\label{sec:Introduction}
 The weak decay of quarks depends on fundamental parameters of the
standard model, including the Cabibbo-Kobayashi-Maskawa (CKM) matrix,
which describes mixing between quark
families~\cite{CKM1,CKM2}. Extraction of these parameters from weak
decays is complicated because the quarks are confined within
color-singlet hadrons as described by quantum chromodynamics (QCD). An
essential tool used in this extraction is the heavy quark expansion
(HQE) technique~\cite{bigishifman}. In HQE the total decay width of a
heavy hadron is expressed as an expansion in inverse powers of the
heavy quark mass $m_q$. At $\mathcal{O}(1/m_b)$ the lifetimes of all
$B$ hadrons are identical. Corrections to this simplification are given by $\mathcal{O}(1/m_b^2)$ and $\mathcal{O}(1/m_b^3)$ calculations leading to
the predicted lifetime hierarchy: $\tau(B^{\pm})>\tau(B^0) \approx
\tau(B^0_s)>\tau(\Lambda_b) \gg \tau(B_c)$ and quantitative predictions
of the lifetime ratios with respect to the $B^0$
meson~\cite{tarantino,Gabbiani1,Gabbiani2,Lenz:2008zz,Beneke:2002rj,Franco:2002fc}.

The Tevatron $p\overline{p}$ collider at $\sqrt{s}=1.96$ TeV has the
energy to produce all $B$ hadron species. The decays of these hadrons
are selected by a variety of successive trigger selection criteria applied at 
three trigger levels. Unique to the CDF II detector is the silicon
vertex trigger (SVT), which selects events based on pairs of tracks displaced 
from the primary interaction point. 
This exploits the long lived nature of $B$ hadrons and collects samples of 
$B$ hadrons in several decay modes, targeting in particular the fully hadronic $B$ decays. Many different measurements of the 
properties of $B$ hadrons have been made using samples selected by this 
trigger, examples of which are given
in Ref.~\cite{bsmixing,bsphiphi,dcpv,bcharmless1,bcharmless2}.

However, this trigger preferentially selects those events in which the
decay time of the $B$ hadron is long. This leads to a biased proper decay time distribution. The conventional approach to correct this bias
has been through the use of a full detector and trigger simulation. An
important source of systematic uncertainty, inherent in this
conventional approach, is how well the simulation represents the
data. A full and accurate simulation of data collected by this trigger is particularly difficult due to the dependence on many variables including particle kinematics, beam-interaction positions, and the instantaneous luminosity. The differences between data and simulation are the dominant systematic uncertainties in the
recent CDF measurement of the $\Lambda_b$ lifetime~\cite{mumford}. These systematic uncertainties will be the limiting factor in obtaining precision measurements of $b$ hadron lifetimes in data samples collected by methods that introduce a time distribution bias. In
this paper we present a new analytical technique for correction of the
bias induced by such a trigger. This technique uses no information from
simulations of the detector or physics processes, and thus incurs none
of the uncertainties intrinsic to the simulation based method.

The technique is presented in a measurement of the $B$$^-$ meson
lifetime using the decay mode \btodpi (charge conjugate decays are
implied throughout). This decay channel is chosen as the lifetime of
the $B^-$ is already well known and the high yield available in this
channel allows a good comparison to the world average. This
measurement demonstrates the ability of this method to reduce the
overall systematic uncertainty on a lifetime measurement. A displaced
track trigger is expected to operate at the LHCb detector, and the
technique of lifetime measurement presented here is applicable to any
data where the method of collection induces a bias in the proper decay time
distribution.

%------------------------------------------------------------

\section{Overview}
\label{sec:Overview}
The simulation-independent method, presented here, for removing the trigger-induced
lifetime bias is based on using a
candidate-by-candidate efficiency function for each $B$ meson candidate. This efficiency function is calculated from 
the event data, without recourse to simulation. 
This approach is based on the observation that for a given set of
decay kinematics of the decay $B^{-} \to D^0 \pi^-$ (i.e the four momenta of
the final state particles and the flight distance of the $D$) the
decay time dependent efficiency function has a simple shape that can easily
be calculated from the measured decay kinematics and the known
decay time dependent cuts. This provides a simple and robust method for
taking into account the effect of the trigger by calculating a
different efficiency function for each candidate and applying it, candidate-by-candidate, in a likelihood fit.  The details of this calculation are
presented in Sec.~\ref{sec:removeBiasForSignal}.

As discussed in Ref.~\cite{Punzi:2004wh}, if a candidate-by-candidate quantity
(here, the efficiency function) enters a fit with a signal and
background component, the probability density function (PDF) for this
quantity needs to be included in the fit, unless it happens to be
identical for both components.  In our case, this constitutes a
significant complication as it requires fitting a distribution of
efficiency functions rather than just numbers. This is
accomplished with an unusual application of the Fisher discriminant method
to translate each efficiency function into a single number, described in
Sec.~\ref{sec:SgBgPDF}.

While we do not use any input from simulation in extracting the $B$ lifetime from the data, we do
use simulated events to test our analysis method and also to evaluate
systematic uncertainties. We use a full {\sc geant3}-based detector
simulation~\cite{GEANT3}, (which includes a trigger simulation),
 as well as a detailed fast simulation for high-statistics
studies. The results of the simulation studies are presented in
Sec.~\ref{sec:validation}, and Sec.~\ref{sec:Systematics}. In Sec.~\ref{sec:FitResults} we show the results of applying the
method to our data, and in Sec.~\ref{sec:Conclusion} we summarize our
conclusions. A brief description of the
relevant components of the CDF detector, in particular, the trigger is given in Sec.~\ref{sec:Detector}, followed by the description of the event
reconstruction, data selection and sample composition in Sec.~\ref{sec:Data}.

%%

%------------------------------------------------------------

\section{The CDF II Detector and Trigger Selection}
\label{sec:Detector}
This analysis uses data corresponding to 1 fb$^{-1}$ of integrated luminosity collected by the CDF II detector at the Fermilab Tevatron using $p \bar{p}$ collisions at
$\sqrt{s}=1.96$ TeV. The data were collected during the first four years
(2002--2006) of the ongoing Run-II data taking period. The CDF II detector is
described in detail elsewhere~\cite{cdf_detector}. A brief
description of the most relevant detector components for this
analysis follows.
\subsection{CDF II Detector}
\label{sec:detector}
The CDF II detector has a cylindrical geometry with forward-backward
symmetry. It includes a tracking system in a 1.4 T magnetic field, coaxial
with the beam. The tracking system is surrounded by calorimeters
and muon detection chambers. A cylindrical coordinate system, $(r,\phi,z)$ is used with origin at the geometric center of the detector, where $r$ is the 
perpendicular distance from the beam,
$\phi$ is the azimuthal angle and the $\hat{z}$ direction is in
the direction of the proton beam. The polar angle $\theta$ with
respect to the proton beam defines the pseudorapidity $\eta$ which is
given by $ \eta = -\ln (\tan \frac{\theta}{2})$.

The CDF II detector tracking system consists of an open cell argon-ethane gas
drift chamber called the central outer tracker (COT)~\cite{COT}, a
silicon vertex microstrip detector (SVX-II)~\cite{SVX}, and an intermediate silicon layer detector (ISL)~\cite{ISL}. The SVX-II is
96 cm long, with three sub-sections in $z$ and has five concentric layers
of double sided silicon microstrip detectors from $r$=2.45 to
$r$=10.60 cm segmented into 12 wedges in $\phi$. The COT is 310 cm
long, consisting of 96 sense wire layers grouped into eight alternating
axial and 2$^{\circ}$ stereo superlayers. The ISL lies between a radius of 20.0 and
29.0 cm and helps in extending the $\eta$ coverage of the SVX-II and
COT. Together the SVX-II, ISL, and COT provide $r$-$\phi$ and $z$
measurements in the pseudorapidity range $\mid \eta \mid <$2
 or $\mid \eta \mid <$1 for tracks traversing all eight COT superlayers.

\subsection{Track Parametrization}
A charged particle has a helical trajectory in a constant magnetic
field. A description of the five parameters used to describe charged particle tracks at the CDF 
experiment follows. In the transverse plane, which is the plane perpendicular to
the beam direction and described by $x$ and $y$ coordinates, the helix is parametrized with track curvature
$C$, impact parameter $d_0$, and azimuthal angle $\phi_0$. The
projection of the track helix onto the transverse plane is a circle of
radius $R$, and the absolute value of the track curvature is $\mid C
\mid = \frac{1}{2R}$. The curvature is related to the magnitude of the
track's transverse momentum, $p_T$, by $\mid C \mid = \frac{1.49898
\cdot 10^{-3} \cdot B}{p_T}$, where $C$ is in cm$^{-1}$, $B$ is in
Tesla and $p_T$ is in GeV/$c$, where $c$ is the speed of light in vacuum. The 
sign of the curvature matches the sign of the track charge. The absolute value of $d_0$ corresponds to the 
distance of closest approach of the track to the beam line. The sign of $d_0$ 
is taken to be that of $(\hat{p} \times \hat{d}) \cdot \hat{z}$, where 
$\hat{p}$ is the  unit vector in the direction of the particle
trajectory, $\hat{d}$ is the direction of the vector from the primary
interaction point to the point of closest approach to the beam, and $\hat{z}$ is the unit vector in the direction of increasing $z$. The angle $\phi_0$ is the azimuthal angle
between $\hat{x}$ and the particle momentum at closest approach. The
two remaining parameters that uniquely define the helix in three
dimensions are the cotangent of the angle $\theta$ between the $z$ axis
and the momentum of the particle and $z_0$, the position along the $z$
axis at the point of closest approach to the beam.

\subsection{Trigger Selection}
\label{sec:trigselec}

The CDF II detector hadronic $B$ trigger is at the heart of this analysis. It collects
large quantities of hadronic $B$ decays, but biases the measured proper decay time distribution through its impact-parameter-based selection. The CDF II detector has a three level trigger system. The first two levels, level 1 (L1) and level 2 (L2), are implemented in hardware and the third, level 3 (L3), is implemented in software on a cluster of computers using reconstruction algorithms similar to those used offline.  The CDF trigger has many different configurations of selection requirements designed to retain specific physics signatures. In this paper we refer to the family of triggers aimed at collecting samples of multi-body hadronic $B$ decays as the ``two track trigger''.

At L1 the trigger uses information from the extremely fast tracker (XFT)~\cite{XFT}. It requires
two tracks in the COT and imposes criteria on track $p_T$ and opening angle. At L2 the silicon vertex trigger (SVT)~\cite{SVT}, which uses silicon hits and fast pattern recognition,
reapplies the $p_T$ criteria, associates silicon hits with each XFT
track and requires that the absolute value of each track's $d_0$ lies
between 120 and 1000~$\mu$m. 

A determination of the beam collision point or primary vertex is
continuously made by the SVT during each data taking period (defining a
run) and is used by all relevant triggers. After data taking is
complete, the offline algorithm uses full detector information and
fully reconstructed three dimensional tracks for a more accurate
determination. At L2 additional criteria are imposed on variables calculated from
each track pair found by the SVT.  The variables are: the product of
the track charges (opposite or same sign), a track fit $\chi^2$ quantity, the opening angle of the two tracks in the transverse plane, the scalar sum of the $p_T$ of the two
tracks, and the $L_{xy}$, where the $L_{xy}$ is the projection of the distance between the primary vertex and two track intersection along the direction of the sum of the two track $\vect{p_t}$. The L3 trigger uses a full reconstruction of the event with all
detector information, (although using a slightly simpler tracking algorithm
than the one used offline) and reconfirms the criteria imposed by
L2. In addition, the difference in $z_0$ of the two tracks is required
to be less than 5 cm removing events where the pair of tracks originate from different collisions within the same crossing of $p$ and $\overline{p}$ bunches. The impact parameter for any given track
measured by the L2 (SVT) is, in general, different from the impact parameter
calculated by the L3 or offline reconstruction algorithms for the same
track due to the differing algorithms. These different measurements of impact parameter are referred
to in this paper as \dzSVT, \dzLt, and \dzOff from L2 (SVT), L3, and
offline algorithms, respectively. 

Three different two-track trigger configurations are used in this analysis. Their criteria are summarized in Table~\ref{secDetector:tab:trigcut} in
terms of the quantities described above. It is clear that the impact
parameter and $L_{xy}$ requirements will preferentially select
long-lived $B$ hadron decays over prompt background. The three selections are referred to as the low-$p_T$, medium-$p_T$ and high-$p_T$ selections. This is a reference to their single track $p_T$ ($>$ 2.0, 2.0, 2.5 GeV/$c$, respectively) and track pair $p_T$ scalar sum ($>$ 4.0, 5.5, 6.5 GeV/$c$, respectively) selection requirements.

The requirements of the three trigger selections mean that any event that passes the
high-$p_T$ selection, simultaneously satisfies the requirements of the low and medium $p_T$ selections. The three separate selection criteria exist because of the need to control the high trigger acceptance rates that
occur at high instantaneous luminosity due to high track multiplicity. The rates are controlled by the application of prescaling, which is the random rejection of a predefined fraction (dependent on the instantaneous luminosity) of events accepted by each trigger selection. Therefore only the higher purity,
but less efficient, high-$p_T$ selection is available to accept events at
higher luminosities.

\begin{table*}
\caption[Trigger selection criteria summary]{Trigger selection criteria for the three two-track trigger selections. We use $n/a$ where no criterion is applied. $\dagger$ - The trigger requirements on the $\chi^2$ were altered during the data taking period. The quantity in brackets refers to the first 0.21 fb$^-1$ collected.\label{secDetector:tab:trigcut}}
\begin{tabular}{lcccc}\hline \hline
\textbf{Trigger criteria L1} &Units& \texttt{Low $p_T$} &
\texttt{Medium $p_T$} & \texttt{High $p_T$} \\ \hline 
Minimum track $p_T$ & GeV/$c$ & 2.0 & 2.0 &2.5 \\  
Two track charge product&- & $n/a$ & $-1$ & $-1$ \\ 
Two track max $\Delta \phi$&degrees & 90$^\circ$ & 135$^\circ$ &
135$^\circ$ \\ 
Minimum two track $p_T$ scalar sum& GeV/$c$ & 4.0 & 5.5 & 6.5 \\ \hline 
\textbf{Trigger criteria L2}& & & & \\ \hline 
Minimum $|\dzSVT|$ & $\mu$m & 120 & 120 &120 \\ 
Maximum $|\dzSVT|$ & $\mu$m & 1000 & 1000 & 1000 \\ 
Minimum track $p_T$& GeV/$c$ & 2.0 & 2.0 & 2.5 \\ 
Maximum track $\chi^2$  &- & 15(25)$\dagger$ & 15(25)$\dagger$ & 15(25)$\dagger$ \\ 
Two track charge product&- & $n/a$ & $-1$ & $-1$ \\ 
Maximum pair $\Delta \phi$&degrees  & 90$^\circ$ & 90$^\circ$ &
90$^\circ$ \\ 
Minimum pair $\Delta \phi$&degrees  & 2$^\circ$ & 2$^\circ$ & 2$^\circ$
\\ 
Minimum two track $p_T$ scalar sum &GeV/$c$ & 4.0 & 5.5 & 6.5 \\ 
Minimum two track $L_{xy}$&$\mu$m & 200 & 200 & 200 \\ \hline
\textbf{Trigger criteria L3} && & & \\ \hline 
Minimum $|\dzLt|$ & $\mu$m & 80 & 80 &80 \\ 
Maximum $|\dzLt|$ & $\mu$m & 1000 & 1000 & 1000 \\
Minimum track $p_T$& GeV/$c$ & 2.0 & 2.0 & 2.5 \\ 
Maximum track $\eta$&- & 1.2 & 1.2 & 1.2 \\ 
Two track charge product&- & $n/a$ & $-1$ & $-1$ \\
Maximum pair $\Delta \phi$& degrees & 90$^{\circ}$ & 90$^{\circ}$ & 90${^\circ}$ \\
Minimum pair $\Delta \phi$& degrees & 2$^{\circ}$ & 2$^{\circ}$ & 2$^{\circ}$ \\
Maximum pair $\Delta z_0$& cm & 5.0 & 5.0 &5.0 \\ 
Minimum two track $p_T$ scalar sum & GeV/$c$ & 4.0 & 5.5 & 6.5 \\ 
Minimum two track $L_{xy}$&$\mu$m & 200 & 200 & 200 \\ \hline \hline
\end{tabular}

\end{table*}
The SVT single track finding efficiency as a function of $\dzOff$, $\eff(\dzOff)$, is an important factor in this analysis. There have been three improvements in the SVT efficiency over the course of the data taking time period used by this analysis due to changes in the pattern recognition algorithm. These have led to three consecutive time periods in which $\eff(\dzOff)$ has improved. These three periods and different resulting efficiencies are incorporated into the analysis as described in Sec.~\ref{sec:validation}.

%------------------------------------------------------------

\section{Data selection and event reconstruction}
\label{sec:Data}

\subsection{Reconstruction of the decay \btodpi}
\label{sec:btodpi}

The reconstruction of the decay \btodpi uses data collected by the two
track trigger described in Sec.~\ref{sec:trigselec}. Standard track
quality selection criteria are applied to all individual tracks: each
track is required to have $p_T>0.4$ GeV$/c$, $\mid \eta \mid<2$, a
minimum of five hits in at least two axial COT super layers, a minimum
of five hits in at least two stereo COT super layers and a minimum of
three silicon hits in the SVX-II $r$-$\phi$ layers.
Candidate $D^0 \to K^- \pi^+$ or $\overline{D^0} \to  K^+ \pi^-$ are searched for first. As no particle identification is used in this analysis, the search for $D^0$($\overline{D^0}$) candidates considers all pairs of oppositely charged tracks which are then assumed to be $K^-$ and $\pi^+ $($\pi^-$ and $K^+$) and assigned the kaon and pion (pion and kaon) masses, respectively. 
The two tracks are then constrained to
 come from a common vertex and the invariant mass ($m_{D^0}$) and
 $p_T(D^0)$ are calculated. Candidates are required to have a mass
 within 0.06 GeV$/c^2$ of the world average $D^0$ mass, 1.8645
 GeV$/c^2$~\cite{PDG}, and $p_T(D^0)>2.4$ GeV$/c$. The $K^-\pi^+$ pair
 is required not to exceed a certain geometric separation in the
 detector. Defining the separation in the $\eta$-$\phi$ plane, in
 terms of the differences in $\eta$ and $\phi$ of the two tracks, as
 $\Delta R = \sqrt{\Delta \eta^2 + \Delta \phi^2}$, we require $\Delta
 R <2$. The separation in $z_0$ of the two tracks is required to be
 $\Delta z_0 <5$ cm. The candidate $D^0$ is then combined with each
 remaining negatively charged track with $p_T >$ 1 GeV$/c$ in the
 event. These are assumed to be pions from the decay $B^- \to D^0
 \pi^-$. The $D^0$ and the $\pi^-$ are constrained to a common vertex
 assumed to be the decay point of the $B^-$ with the $D^0$ mass
 constrained to the world average. The three tracks can be combined to
 measure the invariant mass of the candidate $B^-$, $m_B$. 

 Proper decay time calculations in this paper are made using distances
 measured in the plane transverse to the beam.  The proper decay time
 of the $B^-$, $t$, is given by
\begin{equation}
\label{eq:t_from_Lxy}
t = \frac{L_{xy}}{c \left( \beta \gamma\right)_T} 
  = L_{xy} \cdot \frac{m_B}{c p_T} ,
\end{equation}

 where $L_{xy}$ is the projection of the distance from the primary
 vertex to the $B^-$ vertex along the direction of the transverse
 momentum of the $B^-$ and $(\beta \gamma )_T = \frac{p_T}{m_B}$ is
 the transverse Lorentz factor. The statistical uncertainty on
 $L_{xy}$, $\sigma_{L_{xy}}$, is calculated from the full covariance
 matrix of the vertex constrained fit and is dominated by the primary
 vertex resolution which is approximately 33 $\mu$m. We have used the
 average beam position per run, which is calculated offline for each
 run, as an estimate of the primary vertex position. The uncertainty
 on the proper decay time is calculated by transforming
 $\sigma_{L_{xy}}$ into the $B$ rest frame.

 To reduce background we require that the $B^-$ candidate must have:
 $5.23 < m_B < 5.5$~GeV$/c^2$, 0 $< t<$ 10~ps, $p_T >5.5$~GeV$/c$,
 $L_{xy} >350$~$\mu$m, that the impact
 parameter of the $B$ with respect to the beam spot is smaller than 80~$\mu$m, and that $\sigma_{t}<0.333$~ ps where $\sigma_{t}$ is the decay time uncertainty. We also require that the $\chi^2$ of the vertex constrained fit
 is less than 15, that all tracks have $z_0$ within 5~cm of each
 other, and that $\Delta R( D^0, \pi^-) <2$.

 It is possible to reconstruct candidates where no pair of tracks in
 the final state meet the trigger criteria. The lifetime measurement
 method presented here cannot be used on these candidates, and they are removed by reconfirming the trigger.  We require that
 at least one track pair from each candidate decay pass the L2 and
 L3 trigger selection requirements. The particular L2 and L3 selection
 that the decay must pass depends on which trigger selection accepted
 the event during data taking. In the case where more than one trigger
 selection was satisfied during data taking, we require that the
 candidate satisfies the least stringent selection. Reconfirmation of
 the trigger requires that the offline reconstructed tracks are
 associated to L2 and L3 tracks in the event. To match an offline
 track to a L2 or L3 track we calculate the $\chi^2 =
 \left(\frac{\Delta C}{\sigma_{C}}\right)^2 +
\left(\frac{\Delta \phi }{\sigma_{\phi}}\right)^2$ between an offline track and
each L2 or L3 track in the candidate, where $\Delta C$ and $\Delta \phi$
are the differences between the offline and L2 or L3 track $C$
(curvature) and $\phi$, respectively, and $\sigma_{C}$ and
$\sigma_{\phi}$ are the mean uncertainties on the offline track $C$ and
$\phi$, respectively. The L2 or L3 track that has the lowest $\chi^2$
is associated with the corresponding offline track. If the $\chi^2$ of the L2(L3) track with the lowest $\chi^2$ is greater than 95(25) we consider the match unsuccessful, and deem that the offline track has no L2(L3) matched track. 

Collectively, the trigger selection requirements and
 the cuts made on offline or derived variables are referred to as the
 selection criteria. The kinematics of each track are used to calculate the efficiency function central to this method. We use the following nomenclature to refer to each individual track. The pion originating from the $B^-$
 vertex is referred to as $\pi_B$ and the pion and kaon originating
 from the $D$ vertex are referred to as $\pi_D$ and $K_D$,
 respectively.

\subsection{Sample composition and signal yield}
\label{sec:sampcomp}
\begin{figure*}
\begin{center}
\resizebox{1.0\textwidth}{0.4\textheight}{\includegraphics[clip]{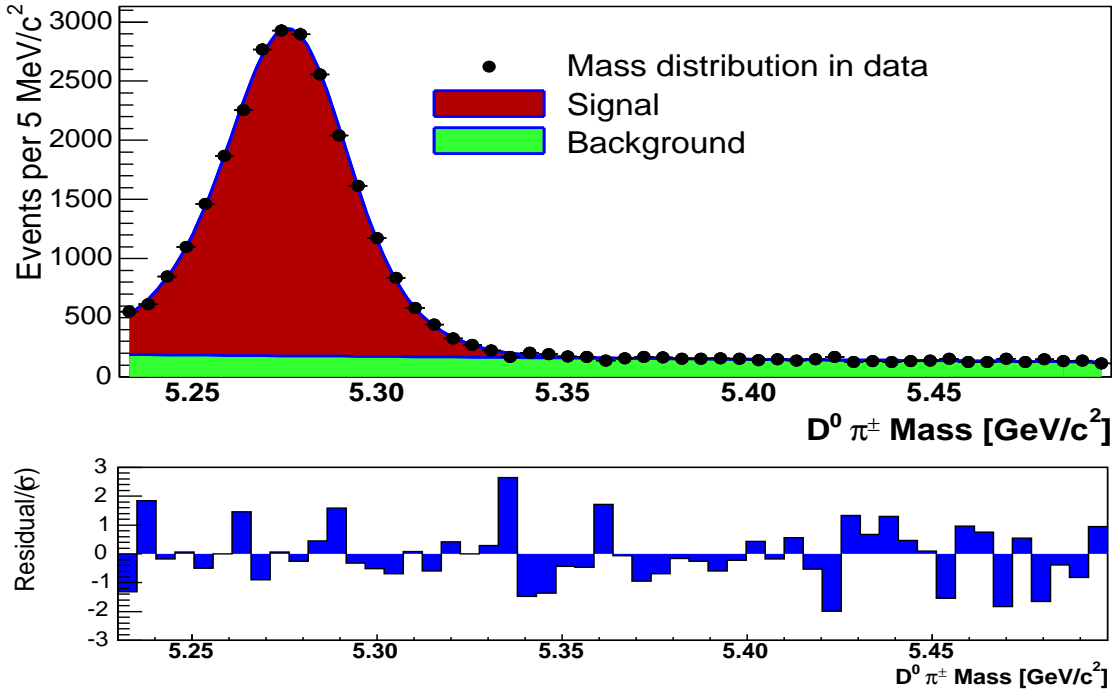}}
\end{center}
\caption[]{The top plot shows the mass fit projection (line) on the data (points). The bottom plot shows the residual divided by the error for each bin: ($N_{\mathrm{fit}}-N_{\mathrm{data}}$)/$\sqrt{N_{\mathrm{data}}}$.}
\label{fig:mass}

\end{figure*}

The invariant $D^0\pi^-$ mass distribution after the selection criteria have been
applied is shown in Fig.~\ref{fig:mass}. The low mass background
sideband and a small part of the signal peak have been removed by the
requirement that $m_B >5.23$~GeV$/c^2$. This cut has been applied to
remove partially reconstructed $B^- \to D^{*0} \pi^-/\rho^-$ and $B^0
\to D^{(*)-} \pi^+/\rho^+$ decays, where only three tracks of the final
state are used in reconstruction leading to a low reconstructed $B$ mass. If
left in the sample, these partially reconstructed $B$ mesons would
bias the proper decay time distribution, since they resemble signal candidates,
but, due to the missing momentum, their proper decay time has been mis-measured
(see \equationref{eq:t_from_Lxy}). Detailed Monte Carlo studies have shown that the applied mass cut leaves the signal peak with a negligible contamination (~0.15$\%$) from partially reconstructed $B^- \to D^{*0}\pi^{-}$ decays. No
other partially reconstructed $B$ hadron decays are expected to
populate this mass range. The Cabibbo suppressed decay \bdk is also
present in this sample, where the kaon from the $B$ is reconstructed as
a pion. The lower mass cut does not remove all of these candidates, but a
tighter cut would remove too many \btodpi candidates. For simplicity, the
\bdk candidates are not fit separately and are treated as \btodpi candidates
for the lifetime determination. This simplification is motivated by
the small size of the contamination (3\%), and the small difference
in reconstructed proper decay time between the $K$ and the $\pi$ mass
assignment of the kaon track which is of order 1$\%$. The resulting systematic uncertainty was evaluated and found
to be negligible (\secref{sec:Systematics}). The mass distribution of
the remaining signal candidates, including both \btodpi and \bdk\, is
modeled by the sum of two Gaussians each with an independent mean and
width. The background candidates are due to track combinations that mimic
the signature of signal decays. The mass distribution of background candidates 
is modeled by a linear function. An alternative description which allows for a second order polynomial to model the background was found to be degenerate with the linear function.

 To determine the signal yield the mass distribution is fit by
 maximizing an unbinned log likelihood, $\mathcal{L}$, which is
 calculated using the mass, $\mass$, for each candidate. The letters $s$
 and $b$ denote whether the PDF describes signal or background
 candidates. The likelihood is given by
\begin{equation}
\begin{split}
\log\left(\mathcal{L}\right) = 
 \log \bigg\{ &
       \prod_i^N \Big[
                       f_s\mathcal{P}\left(\mass|s\right)\\
                          &+ \left(1-f_s\right)\mathcal{P}\left(\mass|b\right)
                \Big]
      \bigg\},\\
\end{split}
\end{equation}
where $f_s$ is the signal fraction and $\mathcal{P}\left(\mass|s\right)$ is given by
\begin{equation}
\begin{split}
\mathcal{P}\left(\mass|s\right) 
= 
   \Big[ &
     \frac{f_1}{{\sigma_1 \sqrt {2\pi } }}
       e^{ - \frac{\left( {\mass - m_1 } \right)^2 }{2\sigma_1^2}}\\
  & + \frac{(1-f_1)}{{\sigma_2 \sqrt {2\pi } }}
       e^{ - \frac{\left( {\mass - m_2 } \right)^2 }{2\sigma_2 ^2 }}
    \Big]
  \cdot \mathcal{A},
\end{split}
\label{e:masspdf}
\end{equation}
where the factor $\mathcal{A}$ is required to satisfy the
normalization condition
\begin{equation}
 \int_{m_{\mathrm{low}}}^{m_{\mathrm{high}}} \mathcal{P}(\mass|s) d\mass =1.
\end{equation} 
$\mathcal{P}(\mass|b)$ is described by a first order polynomial and is given by:
\begin{equation}
\begin{split}
\lefteqn{
\mathcal{P}(\mass|b) =} & \\ & \frac{1-\alpha \mass}{\left[ m_{\mathrm{high}}-m_{\mathrm{low}} -\frac{\alpha}{2}\left(m_{\mathrm{high}}^2-m_{\mathrm{low}}^2\right)\right]} ,
\end{split}
\label{e:massbkgpdf}
\end{equation}
where $m_{\mathrm{low}}$ and $m_{\mathrm{high}}$ are the lower and upper mass limits, 5.23 and 5.5~GeV$/c^2$, respectively.

The free parameters in the mass fit are $m_1$, $m_2$, $\sigma_1$,
$\sigma_2$, $\alpha$, $f_1$, and $f_s$. The data are fit and the mass
fit projection is shown in Fig.~\ref{fig:mass}. From the results of
the mass fit a yield of 23900$\pm$200 signal candidates is determined. We
define the upper sideband to be the candidates with $5.38<m_B<5.5$
GeV$/c^2$. These candidates are retained to constrain the parameters of
the background component of the lifetime fit. The best fit parameters are given in \appref{sec:fittab}.

The results of the mass fit are also used to extract the signal
distribution of various parameters using background subtraction. We
use this technique in several places for cross checks, but not as a method
to extract the lifetime or any other fit parameter. For the purpose of
background subtraction, we define a signal window by $5.25 < m_B <
5.31$ GeV$/c^2$. The results of the mass fit are used to calculate the fraction
of background candidates in the signal region. For any
given parameter, we subtract an appropriately scaled high mass sideband
distribution from the distribution found in the signal region to
obtain the signal distribution in data.

%------------------------------------------------------------

\section{Removing the selection-induced bias for Signal Events}
\label{sec:removeBiasForSignal}

\subsection{Introduction}
\label{sec:4intro}
 In this section we derive the PDF that takes into account lifetime
 bias due to the trigger and other selection criteria without input
 from simulation. Only the case of pure signal is considered in this
 section, whereas the complications introduced by the presence of
 background candidates are discussed in \secref{sec:SgBgPDF}.

 Before describing the PDF in detail, we give a short overview of the
 essential idea behind our method of correcting for the trigger
 effects in a completely data-driven way. We start by considering an unbiased 
 proper decay time distribution, which
 is given by an exponential. To incorporate detector effects, the
 exponential is convolved with a resolution function. For the
 purpose of this measurement, the proper decay time resolution
 function at the CDF detector is adequately described by a single
 Gaussian of fixed width. For a decay with mean lifetime $\tau$ and
 Gaussian proper decay time resolution of width $\sigma_t$, the
 probability density to observe a signal candidate decaying with
 proper time $t_i$, where the subscript $i$ labels the candidate, is
 given by
\begin{equation}
\begin{split}
 \mathcal{P}\left(t_i;\tau|s\right) &= \frac{1}{\tau}e^{\frac{-t_i}{\tau} + \frac{\sigma_t^2}{2\tau^2}}\,\Freq\!\left(\frac{t_i}{\sigma_t}-\frac{\sigma_t}{\tau}\right), \\
\textrm{ where } &
 \Freq\!\left(x\right)= \frac{1}{\sqrt{2\pi}}\int\limits_{-\infty}^{x} e^{\frac{-y^2}{2}} \,dy ,
\label{e:noselec}
\end{split}
\end{equation}
and $s$ indicates that this PDF is for signal events only.
Now consider a dataset subject to the requirement that the lifetime
$t$ is within the interval $t\in [a,b]$. In this case, the PDF in
\equationref{e:noselec} must be modified to take into account this
selection. The effect of the selection can be accounted for by
correct normalization so that the PDF is now
\begin{equation}
%\begin{split}
%\lefteqn{
\mathcal{P}\left(t_i;\tau|s\right)=%} & \\
%& 
\frac{\frac{1}{\tau}e^{\frac{-t_i}{\tau} +
\frac{\sigma_t^2}{2\tau^2}}\Freq\!\left(\frac{t_i}{\sigma_t}-\frac{\sigma_t}{\tau}\right)
}{\int\limits_{a}^{b} \frac{1}{\tau}e^{\frac{-t}{\tau} +
\frac{\sigma_t^2}{2\tau^2}}\Freq\!\left(\frac{t}{\sigma_t}-\frac{\sigma_t}{\tau}\right)dt }.
%\end{split}
\label{eq:effSimpleNorm}
\end{equation}
The same equation can be written as
\begin{equation}
%\begin{split}
%\lefteqn{
\mathcal{P}\left(t_i;\tau|s\right)=%} & \\
%& 
\frac{E(t)|_{t=t_i} \frac{1}{\tau}e^{\frac{-t_i}{\tau} +
\frac{\sigma_t^2}{2\tau^2}}\Freq\!\left(\frac{t_i}{\sigma_t}-\frac{\sigma_t}{\tau}\right)
}{\int\limits_{-\infty}^{\infty} E(t)  \frac{1}{\tau}e^{\frac{-t}{\tau} +
\frac{\sigma_t^2}{2\tau^2}}\Freq\!\left(\frac{t}{\sigma_t}-\frac{\sigma_t}{\tau}\right)dt },
%\end{split}
\label{eq:effSimpleE}
\end{equation}
where, for the example given here, the value of the efficiency
function $E(t)$ is one for $a<t<b$ and zero otherwise. This is essentially the form of the lifetime PDF for candidates collected by the selection criteria at CDF, except that the function
$E(t)$ will take a slightly more complicated form, and will be
different candidate by candidate. We indicate this by adding a subscript $i$
that labels the candidate, \acc\ . The introduction of $\varepsilon_s$ is made because the efficiency function will also be shown to depend on $\varepsilon_s$ which is the single track finding efficiency at level 2. This candidate-by-candidate efficiency function \acc\ is
the crux of this analysis, and it will be described in detail in the
following sections. 

The CDF trigger selects on the impact parameters of the tracks in the
decay. The impact parameter requirements can be translated to an upper and lower
decay time selection for each candidate. These upper and lower lifetime limits
depend on the kinematics of the decay and therefore differ for each candidate $-$ hence the need for a candidate-by-candidate \acc. 

In order to calculate the efficiency function, \acc\, for a given candidate we require: the individual candidate's decay kinematics, measured in the data; the \emph{single track} finding efficiency \effSVT (also extracted from the data); and the trigger and offline criteria, collectively referred to by the symbol \trig. In terms of these variables, the PDF for a candidate with decay
time $t_i$ is
\begin{equation}
\begin{split}
\lefteqn{\mathcal{P}\left(t_i;\tau|\trig,\acc,s\right)=} & \\
& 
\frac{\acc|_{t=t_i}\times\frac{1}{\tau}e^{\frac{-t_i}{\tau} +
\frac{\sigma_t^2}{2\tau^2}}\Freq\!\left(\frac{t_i}{\sigma_t}-\frac{\sigma_t}{\tau}\right)
}{\int\limits_{-\infty}^{\infty}\acc\times \frac{1}{\tau}e^{\frac{-t}{\tau} +
\frac{\sigma_t^2}{2\tau^2}}\Freq\!\left(\frac{t}{\sigma_t}-\frac{\sigma_t}{\tau}\right)dt }.
\end{split}
\label{eq:effunbiased}
\end{equation}

To summarize, we use a different efficiency function $\acc$ for each
candidate~$i$, which ensures the correct normalization of the lifetime
PDF given the selection. We calculate each $\acc$ analytically from
the candidate's decay kinematics and the selection criteria, in a
completely data-driven way, without recourse to Monte Carlo. The exact
form of \acc, and how it is calculated, is discussed next.

\subsection{Calculation of \acc}
\label{s:calceff}
\subsubsection{Scanning through different potential proper decay times}
In order to find the function $\acc$ for a given candidate $i$, we need to
find the trigger efficiency for that candidate for all possible $B$ proper
decay times. We scan through different $B$ decay times by translating
the $B$ decay vertex along the $B$ flight direction, defined by the
reconstructed $B$ momentum.
 At each point in the scan, we re-calculate all decay-time dependent
 properties of the candidate, in particular the impact parameters and decay
 distance. Properties that are independent of proper decay time (before selection is
 applied), such as the four momenta of all particles or the flight
 distance of the intermediate $D$ meson, remain constant.
 We re-apply the trigger and other selection criteria to the translated
 candidate. If the translated candidate fails the selection criteria, $\acc$ is
 zero for that candidate at the corresponding decay time. Otherwise $\acc$
 is non-zero at time $t$ and its exact value
 depends on the SVT (L2) track-finding efficiency, \effSVT. This method of scanning through different potential proper decay times allows for the determination of the effective upper and lower decay time cuts applied by the selection criteria.
 This process is illustrated and described in detail in \secref{sec:accExample}. Prior to this, we discuss two complications to the basic idea presented
 above. 
The SVT has a track finding efficiency smaller than that of offline track finding efficiency. The SVT track finding efficiency varies as a function of the track impact parameter. The impact of this variation and the necessary changes to the basic idea are discussed in \secref{varSVTeff}. A secondary complication is that at different stages in the event reconstruction and selection, different algorithms are used to
 calculate the track parameters - very fast algorithms at L2,
 more detailed ones at L3, and finally the full tracking and vertexing
 in the final offline reconstruction. The measured values of track
 parameters such as impact parameters differ slightly depending on the
 algorithm used for the calculation. Section ~\ref{sec:translating} describes how the different measurements of impact parameter are accounted for.

\subsubsection{The value of \acc\ and its dependence on the SVT
 track finding efficiency}
\label{varSVTeff}
\paragraph{The need to include the dependence on \effSVT}
If the track-finding efficiency is independent of proper decay time, one can base a fit on a PDF \emph{given} that a certain track combination has been
reconstructed and seen by the trigger. This would imply that the track finding efficiency is constant as a function of the impact parameter since the decay time and the impact parameter are correlated. In the case where the track-finding efficiency is proper decay time independent, the set of tracks seen by
the trigger would be treated exactly in the same way as the decay
kinematics, i.e. as something that can be kept constant as the decay distance
is changed for the efficiency function evaluation. Given that a
certain track combination has been found, the trigger efficiency at a
certain decay time is either $1$ (passes selection) or $0$ (fails),
independent of \effSVT. This PDF would ignore one factor: the probability
that exactly this track combination has been found. If this factor is
proper decay time independent it does not affect the maximum of the likelihood and
hence the result of the fit. 

The level 3 tracking algorithms are very similar to those used offline and the level 3 track finding efficiency as a function of offline impact parameter is constant. Therefore the track finding efficiency at level 3 is decay time-independent and the situation is that described above; the level 3 trigger efficiency is a time independent constant for all decay times that pass the selection criteria. Therefore it is not necessary to consider the effect of the level 3 track finding efficiency further. However, the situation at level 2 is more complicated.

\begin{figure}[htbp]
\begin{center}
\includegraphics[width=0.99\columnwidth]{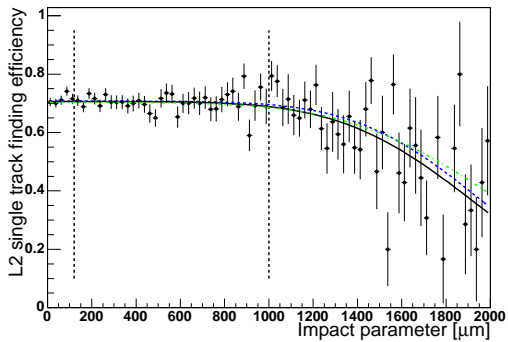}
\end{center}
\caption[Single Track Finding efficiency as a function of impact
parameter]{The L2 single track finding efficiency, relative to the
  offline efficiency, as a function of $|\dzOff|$. The points represent
  the data. The vertical dashed lines represent the trigger selection
  requirements. The fitted curves represent possible descriptions of the
  efficiency which are described in the discussion of systematic
  uncertainties in \secref{s:svtd0}. }
\label{f:svtnonflat}

\end{figure}

Figure~\ref{f:svtnonflat} shows the SVT track
finding efficiency for tracks found in the offline reconstruction, in data, as a function of the track's offline impact parameter $|\dzOff|$. Figure~\ref{f:svtnonflat} shows that the SVT track finding efficiency of the CDF II detector depends on the track impact parameter, and therefore on the decay time of the parent particle. The SVT track finding efficiency is approximately constant for $0<|\dzOff|<
\un{1000}{\mu m}$ and falls rapidly for \un{|\dzOff| > 1000}{\mu m}. The efficiency distribution is obtained from the signal region of the data sample used in the fit, using the following method: the efficiency prior to triggering is obtained by considering the sub-sample of candidates where two particular tracks can pass the trigger requirements. For these candidates, the remaining third track is used to obtain the SVT track finding efficiency.

Even though \effSVT\ is approximately constant within the trigger
acceptance requirements, the rapid drop after \un{|\dzOff| > 1000}{\mu}m,
introduces a particular problem. The trigger efficiency is calculated
depending on which tracks are found by the SVT. If \effSVT\ is
constant for all impact parameters, then the tracks which \emph{were
actually} found by the SVT can be used to calculate the trigger
efficiency, and we can assume that the \emph{same} tracks would be
found as the decay vertex is scanned along the direction of the B
momentum. However, since \effSVT\ does vary with $|\dzOff|$, the
probability of the SVT finding tracks is dependent on the decay
position. The more track combinations there are that pass the
selection criteria, the higher the probability that at least one is
found by the SVT. Under these circumstances, the proper decay time
dependence of the SVT trigger efficiency has to be taken into account,
which requires some parametrization of the single track finding
efficiency as a function of \dzOff.

\paragraph{Parametrizing $\effSVT(|\dzOff|)$}
 While the inclusion of the single track finding efficiency in the PDF
 increases the complexity of the measurement, we can take, as a good
 approximation, the following simple model.  We model the SVT track
 finding efficiency as constant for $0<|\dzOff|< \un{1000}{\mu m}$. We
 treat all tracks with $|\dzSVT| > \un{1000}{\mu m}$ as not-found by the
 SVT (which does not affect the trigger decision as it requires
 $\un{120}{\mu m}<|\dzSVT|< \un{1000}{\mu m}$), so that we can describe
 the SVT efficiency by the following simple description:
\begin{equation}
 \effSVT(\dzOff) = \left\{
 \begin{array}{rl}
   \effSVT &\mathrm{if}\; |\dzOff| < \un{1}{mm} \\
      0    &\mathrm{otherwise}
 \end{array}
 \right\}.
\end{equation}
The value of \effSVT\ is determined simultaneously with the lifetime and other parameters in the fit to data and not from \figref{f:svtnonflat}.
 The consequence on the lifetime measurement of the small deviations
 of the real SVT efficiency from this simple model are discussed in 
 Sec.~\ref{s:svtd0}. We also assume that there is no variation in
 track finding efficiency as a function of track $p_T$ or $\eta$. Such
 variations can alter the probability of finding a particular track
 combination. However as these are time independent, the effect on the
 lifetime measurement is expected to be small. This is also discussed
 in \secref{s:svtd0}, where we show that the effects of these
 simplifications on the lifetime measurements are indeed, sufficiently
 small. There is an alternative, simpler approach, that does not
 depend on \effSVT, which is suitable in situations where the
 track-finding efficiency is constant over a larger range than for the
 SVT at the CDF II detector. This is discussed in
 \appref{sec:simplerPdf}.

\paragraph{Calculating \acc}
The value of \acc\ for a given decay time is the probability that at
least one of the possible track combinations that pass the trigger
criteria is in fact found by the L2 tracking algorithms. For example,
if there is only one track pair in the candidate that can pass the selection
requirements, then the probability of finding both those tracks is
$\effSVT^2$, where we simply take the product of two single track finding efficiencies. For a three body final state, where there are two
possible track pairs that pass the trigger, the probability is given by $2\effSVT^2 -
\effSVT^3$. In cases where there are three possible track pairs (only possible for the \texttt{low}-$p_T$ selection that makes no requirement on track charge), the probability to find sufficient tracks to pass the trigger
is $3\effSVT^2-2\effSVT^3$.

\subsubsection{Translating Online and Offline quantities}
\label{sec:translating}
To calculate the trigger efficiency for all possible $B$ proper decay times we scan through different $B$ decay points along the $B$ flight path and determine the probability that the trigger was passed at that point. As we re-apply the trigger selection, we always base the decision on the
 quantities accessible to the relevant trigger level i.e., L2
 criteria to SVT tracks, L3 criteria to L3 tracks, and offline criteria
 to the fully reconstructed offline tracks. Certain quantities such as the track momentum or the opening angle between two tracks are decay time independent and will remain constant as the vertex is translated along the $B$ flight path. Other quantities such as the impact parameter will change. Therefore, as we translate
 the $B$ decay along its flight direction, we need to re-calculate the
 decay time dependent quantities for each level: L2, L3, and offline.

It is trivial to calculate the offline impact parameters and reconstructed proper decay time as the candidate is translated along its flight path. Furthermore, as \acc\ is a function of the \emph{offline-reconstructed} proper decay time, rather than the true decay time, it is not necessary to reconsider the effects of detector resolution. This means that there is a simple, one-to-one relationship between the offline-reconstructed decay time of the translated candidates and the
 other time-dependent offline quantities such as impact parameters and
 $L_{xy}$, without the need to take into account further resolution effects. We aim to retain a similarly simple direct relationship between proper decay
 time and trigger cuts for the online quantities as well. Since all L2 and L3
 decay time dependent quantities ($d_0$, $L_{xy}$) are calculated from the
 impact parameters of the tracks, the value of the online $d_0$ is the only parameter we need to
 consider.

\begin{figure}[htbp]
  \subfigure[]
  {\includegraphics[scale=0.35]{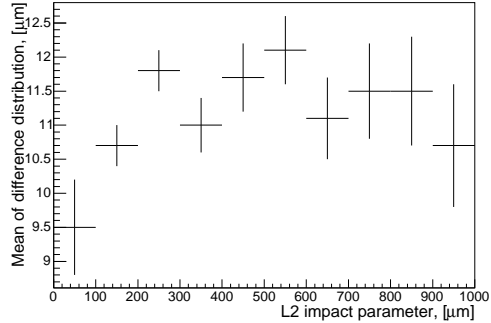}
\label{f:means}}
\qquad
\subfigure[]
{\includegraphics[scale=0.35]{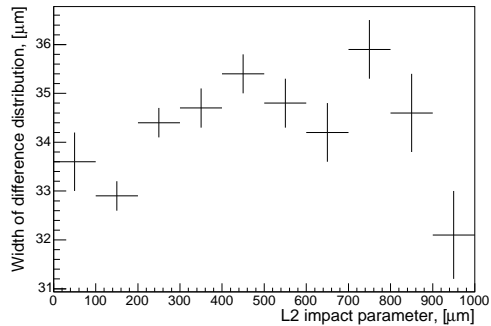}
\label{f:widths}}
\caption[]{The difference, $(\Delta d_0)_{L2}$ is binned as a function of $|\dzSVT|$ and fitted to a Gaussian. The mean of the fitted Gaussian is shown in (a) while the width is given in (b). The variation is of order a few microns.\label{fig:meanswidths}}
\end{figure}

 As we translate the candidate along the $B$ flight path, we re-calculate each track's online
 $d_0$ at L2 and L3 ($d_0^{L2}$ and $d_0^{L3}$), by assuming that the differences between online
 and offline quantities are not decay time dependent. This way, we can treat
 this difference in exactly the same way as the other proper decay time independent
 quantities in the candidate, such as track $p_T$. We measure the differences in each
 candidate and keep them constant as we translate the candidate along the $B$ flight path. The difference between the L2 and offline impact parameter, $(\Delta
 d_0)_{L2}= d_0^{L2} - \dzOff$, could vary as
 a function of impact parameter due to the finite hit recognition
 patterns used to measure the L2 impact parameter. We verify in data
 that $(\Delta d_0)_{L2}$ is time independent. To check this, we
 calculate $(\Delta d_0)_{L2}$ and bin it according to track $|\dzSVT|$. In each bin, the $(\Delta d_0)_{L2}$  distribution
 is fitted with a Gaussian, and the mean and width of the
 fitted Gaussian for different impact parameter ranges is shown in
 Fig.~\ref{fig:meanswidths}. There are some deviations from a straight
 line, but there is no systematic dependence on impact parameter, and
 hence on impact parameter resolution as a function of decay time. Variations in the impact parameter resolution, such as those observed in data, could lead to a bias on a lifetime measurement. This is addressed in \secref{sec:Systematics} and we find any systematic uncertainty on the lifetime due to this variation to be very small (\un{0.02}{ps}). 

 The $(\Delta d_0)_{L2}$ for a given track is measured at the actual
 point of decay by accessing the information of the L2 track that was matched
 to the offline track. This is then used to calculate the translated
 L2 impact parameter $d_0^{L2}(t)$ from the translated offline
 $d_0(t)$ at each point: $d_0^{L2}(t) = \dzOff(t) + (\Delta d_0)_{L2}$. A
 complication arises for those tracks not found by the SVT (such as
 those with $|d_0| \gg$ 1 mm). In this case, a value of $(\Delta
 d_0)_{L2}$ is assigned by drawing a value at random from the
 distribution of $(\Delta d_0)_{L2}$ from tracks where it is possible
 to calculate $(\Delta d_0)_{L2}$. One further issue to consider is
 that the L2 algorithm measures impact parameters to the closest 10
 $\mu$m. To emulate this feature of the L2 tracking algorithm the
 calculated $d_0^{L2}(t)$ is rounded to the closest multiple of 10
 $\mu$m. The same procedure is applied to estimate \dzLt except that
 no discretization is necessary. The online $L_{xy}$ values at L2 and
 L3 for each track pair are then re-calculated from the translated
 L2 and L3 impact parameters of each track.

\subsubsection{Example}
\label{sec:accExample}
\begin{figure*}[htbp]
  \subfigure[nooneline][If the decay occurs at point $a_1$, only one track has an
  impact parameter within the trigger range (shaded region). $\acc$ at
  the corresponding $t$ is zero.]
  {\includegraphics[width=0.43\textwidth]{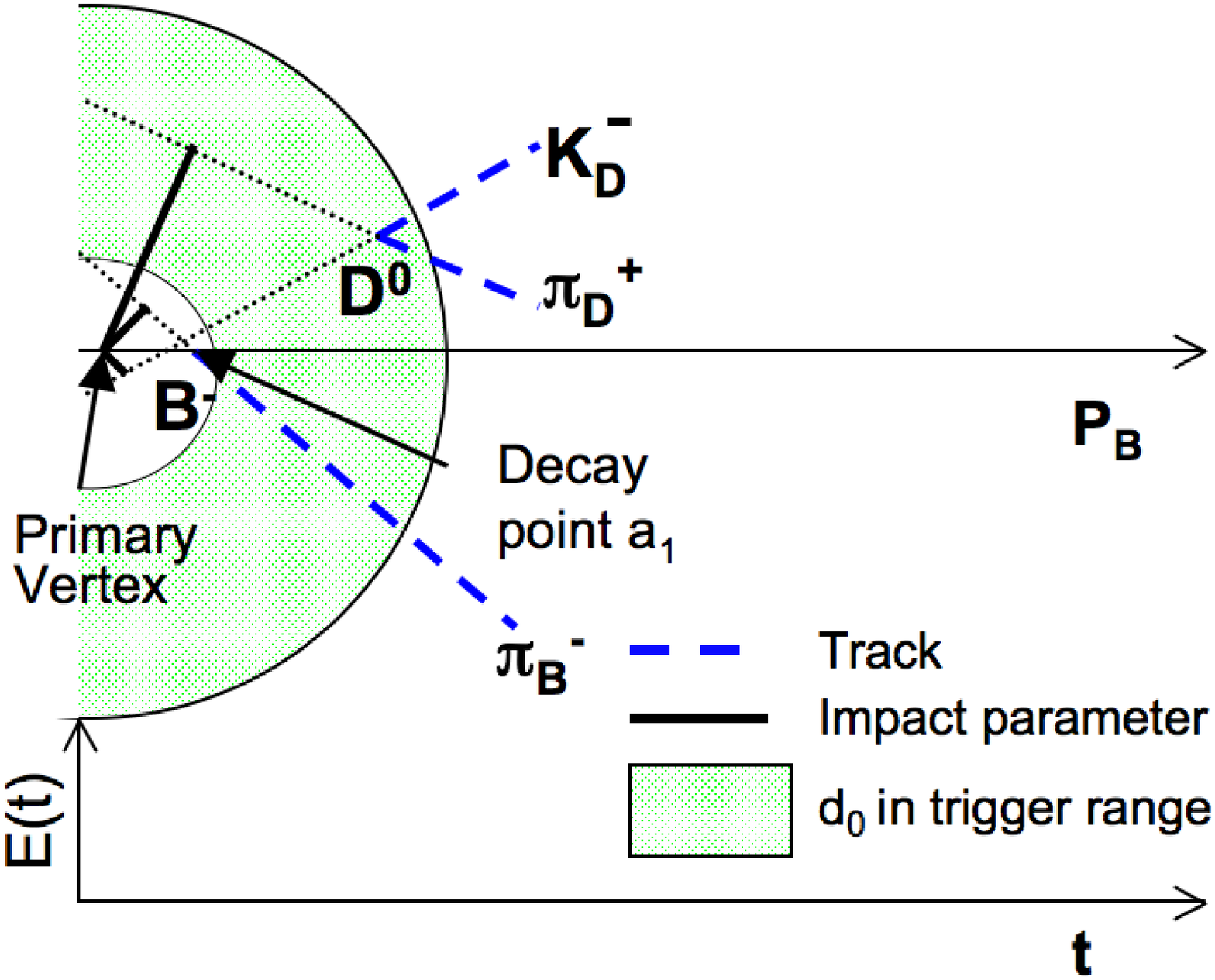}
\label{f:newtrans1}}
%\qquad
%
\subfigure[nooneline][If the decay occurs at $a_2$ there are two tracks which
would pass all trigger and other selection requirements. \acc\ at the
corresponding proper time is the probability that both these tracks
are found by the L2 algorithms. ]
{\includegraphics[width=0.43\textwidth]{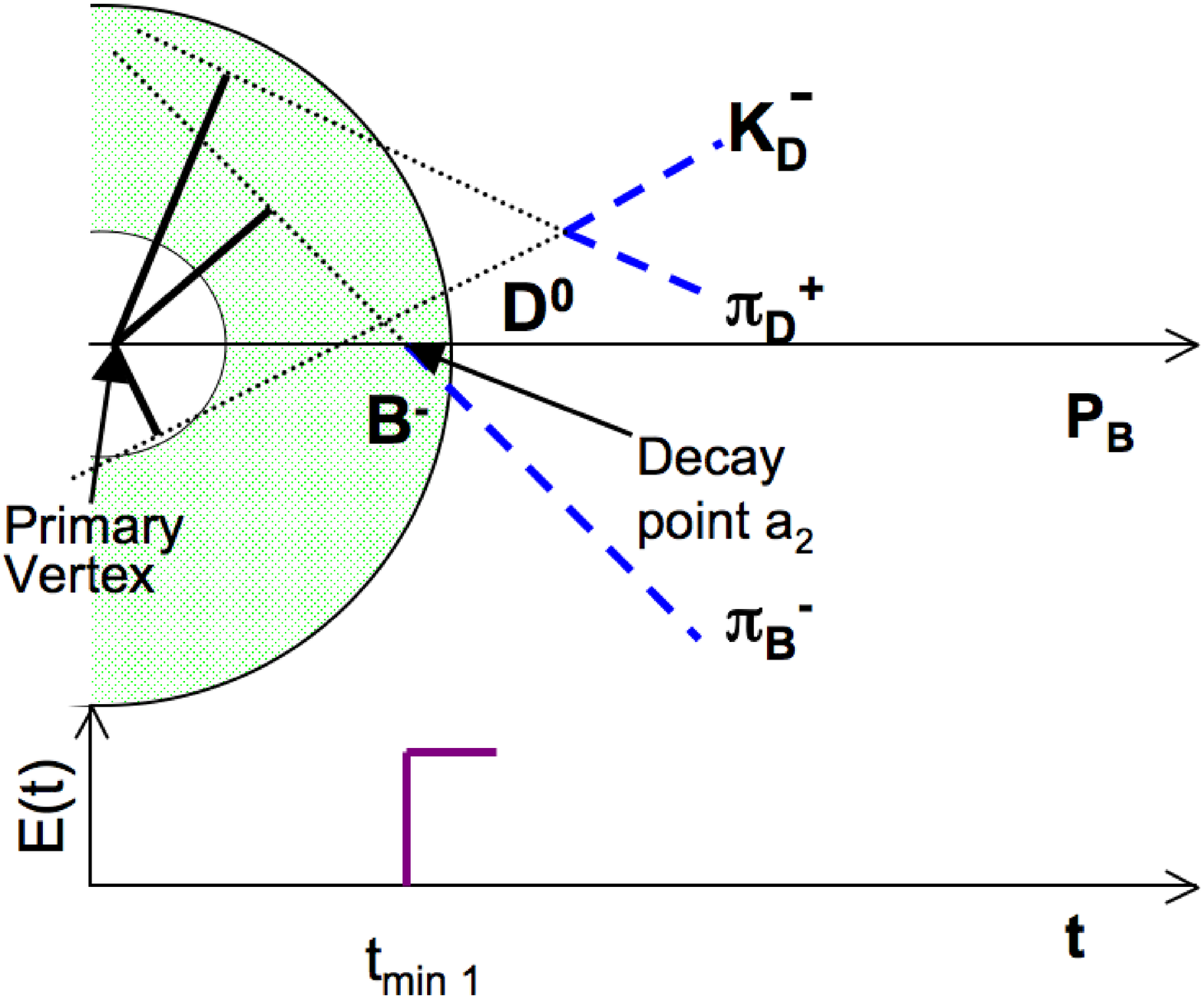}
\label{f:newtrans2}}
%\qquad
%
\subfigure[nooneline][If the decay occurs at $a_3$ there are two tracks pairs
which would pass all trigger requirements. \acc\ rises at the point
$t_{min\ 2}$ as the probability to find at least one of the two
available track pairs is greater than the probability to find a
particular track pair. ]
{\includegraphics[width=0.43\textwidth]{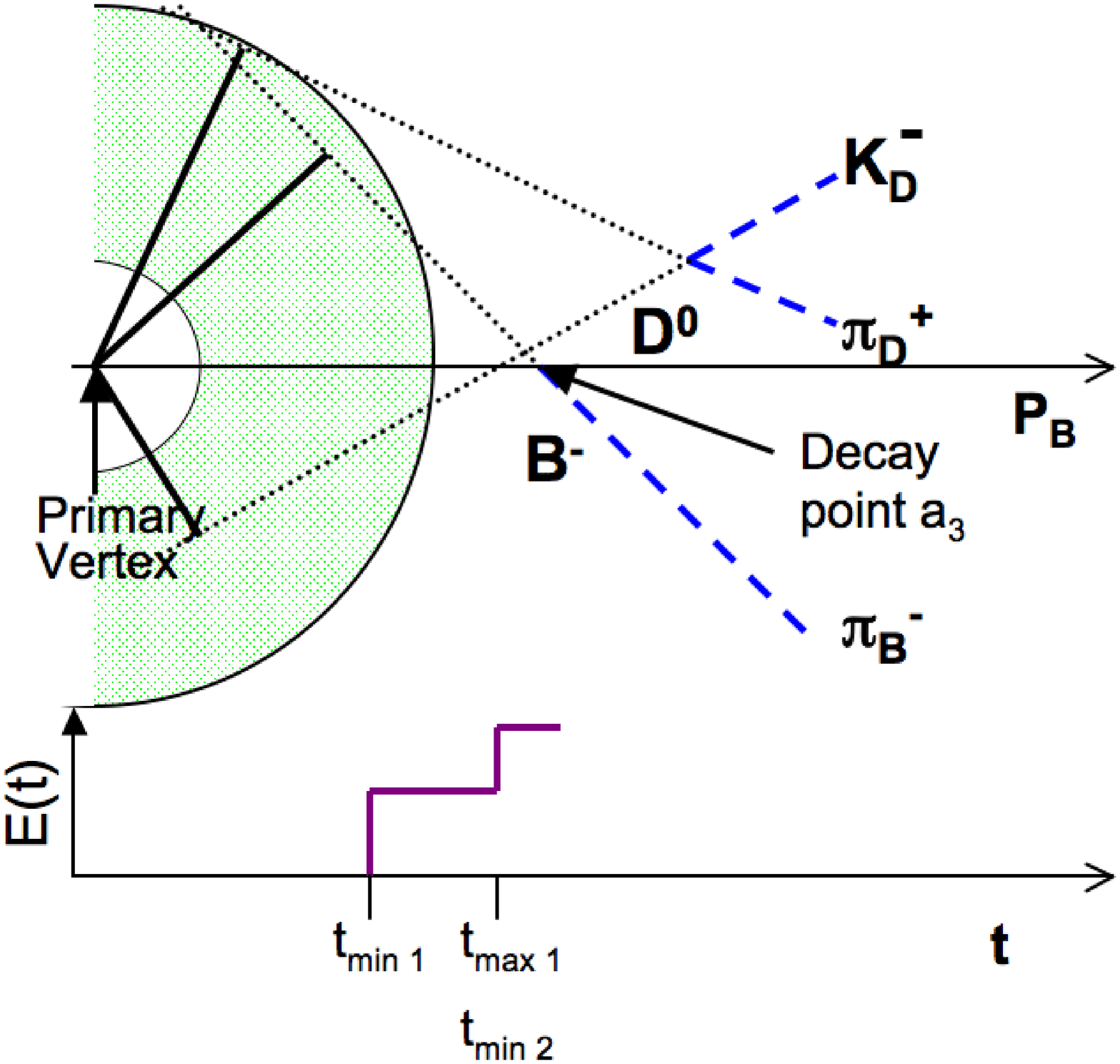}
\label{f:newtrans3}}
%
%\qquad 
\subfigure[nooneline][If the decay occurs at $a_4$ all the track impact
parameters are above the trigger threshold. \acc\ at $t_{max\ 3}$
returns to zero. ]
{\includegraphics[width=0.43\textwidth]{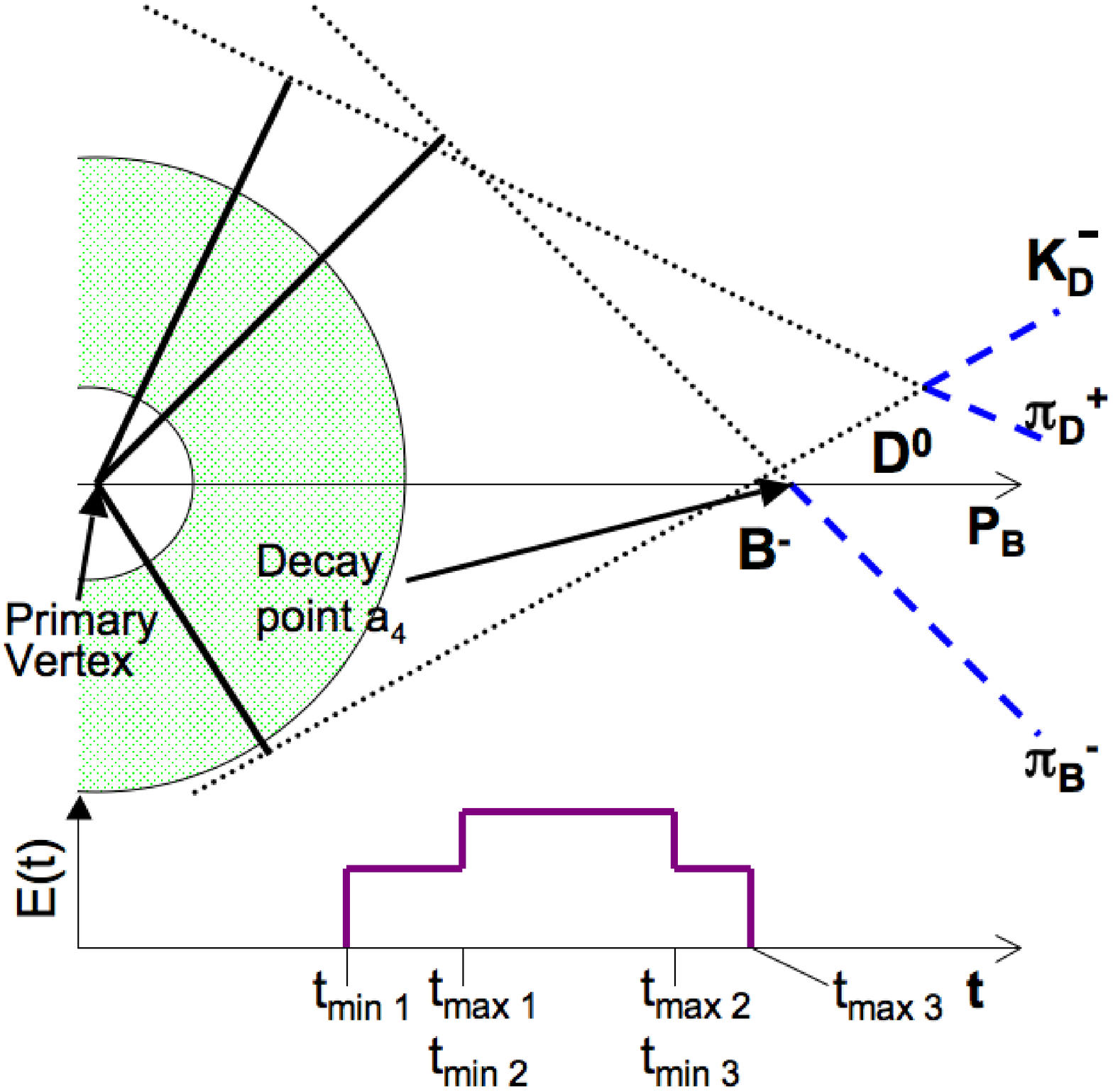}
\label{f:newtrans4}}
\caption[]{
 The decay vertex is translated along the direction of the $B$
 momentum while the decay kinematics are held fixed. At each decay
 point it is determined whether or not the selection criteria could be
 satisfied and \acc\ is calculated. Diagrams not to
 scale.\label{fig:finalacc}}
\end{figure*}

To illustrate the entire process, we describe in detail a specific
example shown in \figref{fig:finalacc} which depicts 
the same decay at four different decay times. For the purposes of this
illustration we assume this decay has been accepted by the
medium-$p_T$ trigger selection.
First, we consider the decay vertex translated to point $a_1$ as shown
in Fig.~\ref{f:newtrans1}. The decay vertex is close to the primary
interaction point and only one track has $|d_0^{L2}|> \un{120}{\mu
m}$, therefore the selection requirements are not met. The value of
\acc\ at the proper decay time corresponding to $a_1$ is $H_1 = 0$, where $H$ is a polynomial function of \effSVT that gives the value of the efficiency function at a given decay position.

In Fig.~\ref{f:newtrans2} the decay vertex has been translated further
along the $B$ momentum direction and is at the point where one track
pair satisfies the trigger selection and the $B$ decay satisfies all other
selection requirements listed in Sec.~\ref{sec:btodpi}. At this
point, $a_2$, the value of \acc\ is given by probability of finding both
the $\pi_B$ and the $\pi_D$ track, which is $H_2 = \effSVT^2$.

As this candidate is further translated along its $B$ momentum direction
it moves into the region where all three tracks can participate in the
trigger decision. In Fig.~\ref{f:newtrans3}, two track combinations
fulfill the trigger requirements, ($\pi_B$, $\pi_D$) and ($\pi_D$,
$K_D$). The remaining combination, ($\pi_B$, $K_D$), does not pass the
trigger in this case as it does not satisfy the opposite charge
requirement of the medium-$p_T$ trigger. The value of \acc\ at the
decay point $a_3$ is the probability that at least one of the two
possible track combinations is found by the SVT, $H_3 = 2\effSVT^2 -
\effSVT^3$.

 In Fig.~\ref{f:newtrans4} the decay vertex has been translated to
 the point $a_4$ where the track impact parameter requirements are not
 satisfied. The value of \acc\ returns to zero at the point where the trigger requirements are not met. Hence \acc\ can
 be described by a series of intervals limited by $\tmin$ and $\tmax$
 and within an interval the value of \acc\ is given by a polynomial in
 terms of $\effSVT$, $H(\effSVT)$. The efficiency function can be
 written in terms of the Heaviside step function $\theta$ as
\begin{equation}
\label{eq:efficiencyFinal}
\begin{split}
\acc =
 \intervSum\!\!\Big\{ & H_{k_i}(\effSVT) \big[
                 \theta\left(t-t_{\mathrm{min}\;{k_i}}\right) 
              \\ & - \theta\left(t-t_{\mathrm{max}\;{k_i}}\right)
            \big]
\Big\}.
\end{split}
\end{equation}

\subsection{The signal PDF and its parameters}
Substitution of \acc\ as given in \equationref{eq:efficiencyFinal} into \equationref{eq:effunbiased} leads to
the following PDF for observing a decay at time $t_i$:
\begin{widetext}
\begin{equation}
\label{eq:withSVTProb}
\mathcal{P}(t_i;\tau|\trig,\acc,s)=
   \frac{  \acc|_{t=t_i}
           \frac{1}{\tau}e^{\frac{-t_i}{\tau} + \frac{1}{2} \frac{\sigma_t^2}{\tau^2}}
           \Freq\left( \frac{t_i}{\sigma_t} - \frac{\sigma_t}{\tau}\right)
   }{
   \intervSum
    \mathrm H_{k_i}(\effSVT)
    \left[
           - e^{\frac{-t}{\tau} + \frac{1}{2} \frac{\sigma_t^2}{\tau^2}}
           \Freq\left( \frac{t}{\sigma_t} - \frac{\sigma_t}{\tau}\right)
           + \Freq\left(\frac{t}{\sigma_t}\right)
    \right]_{t=t_{\mathrm{min\; {k_i}}}}^{t=t_{\mathrm{max\; {k_i}}}}
   } .
\end{equation}
\end{widetext}
 We describe the decay time resolution of the detector as a Gaussian with width
 $\sigma_t= 0.087$~ps. This is the average of the calculated candidate-by-candidate
 $\sigma_{t_i}$ of the background subtracted signal region in
 data. Using a single Gaussian based on a single, global $\sigma_t$,
 instead of a candidate-by-candidate value, significantly simplifies the
 analysis and is justified since the PDF is not very sensitive to the
 exact value of $\sigma_t$. This is the case for two reasons:
 the lifetime to be measured, $\mathcal{O}\left(\un{1.6}{ps}\right)$, is much
 larger than $\sigma_t = \un{0.087}{ps}$; and the selection requirements remove the majority of candidates with low decay times.

 In terms of the PDF in \equationref{eq:withSVTProb}, this implies
 that all terms containing $\sigma_t$ only have a small effect on the PDF because $t/\tau
 \gg \half \sigma^2_t/\tau^2$ and
 $F\!\left(\frac{t}{\sigma_t}-\frac{\sigma_t}{\tau}\right)\approx
 F\!\left(\frac{t}{\sigma_t}\right)\approx 1$. These approximations are not made in the PDF, but they illustrate why the
 dependence on $\sigma_t$ is small. In \secref{sec:Systematics} we
 confirm that the systematic uncertainty due to the resolution
 parametrization is small.

 To use this PDF to extract the lifetime, knowledge of \effSVT\ is also
 required. Although \equationref{eq:withSVTProb} could be used to
 simultaneously fit $\tau$ and $\effSVT$, there is extra information
 available in the data that can be used to help determine \effSVT\ with
 greater precision. The extra information used is simply the knowledge
 of exactly which tracks do, and do not, have L2 information. To add this information to the PDF, we introduce a candidate observable
 called track configuration, $\trk$. This observable is defined both by
 $n$, the number of tracks that are within the reach of the SVT ($p_T
 > \un{2.0}{ \mathrm{GeV}}/c$, $|\dzOff| \in \un{[0, 1]}{mm}$), and by $r$, the
 number of those that have L2 information. The configuration also
 distinguishes which specific tracks have L2 information. The
 probability of observing a particular \trk, i.e., that of $n$ tracks
 within the reach of the SVT, a specific set of $r$ tracks have
 matches, while the remaining $n-r$ tracks do not, is given by
\begin{equation}
\label{eq:ptrack}
\mathcal{P}(\trk|\trig, \acc,\tm, s) = \frac{\effSVT^r(1-\effSVT)^{(n-r)}}{\acc|_{t=t_i}},
\end{equation}
 where the factor $\acc |_{t=t_i}$ provides the correct normalisation
 as it is the sum of all possible configurations that could have
 passed the trigger.

 We multiply the probabilities defined in Eqs.~(\ref{eq:withSVTProb})
 and (\ref{eq:ptrack}) to obtain the PDF which is used to
 simultaneously fit the proper decay time and \effSVT. It is given by
\begin{widetext}
\begin{eqnarray}
\label{eq:withSVTProb2}
\begin{split}
\mathcal{P}(t_i;\tau|\trig,\acc,s)\lefteqn{\cdot \mathcal{P}
 (\trk|\trig,\acc,\tm,s) =}& 
\\ & \frac{  \effSVT^r(1-\effSVT)^{(n-r)}
           \frac{1}{\tau}e^{\frac{-t_i}{\tau} + \frac{1}{2} \frac{\sigma_t^2}{\tau^2}}
           \Freq\left( \frac{t_i}{\sigma_t} - \frac{\sigma_t}{\tau}\right)
   }{
   \intervSum
    \mathrm H_{k_i}(\effSVT)
    \left[
           - e^{\frac{-t}{\tau} + \frac{1}{2} \frac{\sigma_t^2}{\tau^2}}
           \Freq\left( \frac{t}{\sigma_t} - \frac{\sigma_t}{\tau}\right)
           + \Freq\left(\frac{t}{\sigma_t}\right)
    \right]_{t=t_{\mathrm{min\; {k_i}}}}^{t=t_{\mathrm{max\; {k_i}}}}
   }.
\end{split}
\end{eqnarray}
\end{widetext}
 In the case of a two body decay, we would always find, in both the
 numerator and denominator of the expression, that
 $\mathrm{H}_{k_i}(\effSVT) = \effSVT^r(1-\effSVT)^{n-r}=\effSVT^2$;
 all factors containing $\effSVT$ would cancel and we would recover
 the expression for two-body decays derived in
 Ref.~\cite{Rademacker:2005ay}.  If there is no upper impact parameter
 cut or equivalent ($t_{\mathrm{max}}=\infty$), and the lower cut is hard enough so that for each candidate $t_{\mathrm{min}} \gg
 \sigma_t$, \equationref{eq:withSVTProb2} reduces to $\frac{1}{\tau}
 e^{-\left(t-t_{\mathrm{min}}\right)/\tau}$, equivalent to a
 re-definition of $t=0$, as used by DELPHI
 in Ref.~\cite{Adam:1995mb}. Other special cases leading to some
 simplifications are discussed in \appref{sec:simplerPdf}.  However,
 none of these apply here and we use the full expression given
 in~\equationref{eq:withSVTProb2}.

%------------------------------------------------------------

\section{Validation of the Method}
\label{sec:validation}

We test the signal PDF derived in \secref{sec:removeBiasForSignal}, and the full PDF with both
signal and background component that will be derived in
\secref{sec:SgBgPDF}, on simulated events. We use two kinds of
simulations: a full \geantIII-based~\cite{GEANT3} detector simulation and a fast
parametric simulation for high statistics studies.
\subsection{The Full Detector Simulation}
\label{sec:FullSim}
 We use the full CDF II detector simulation to test whether the signal
 PDF constructed in \secref{sec:removeBiasForSignal} can correctly
 remove the selection bias. The simulated data samples used for this
 test consist of single $B$ hadrons generated with $p_T$ spectra
 consistent with NLO QCD~\cite{BGen1,BGen2} and decayed with {\sc
 EvtGen}~\cite{EvtGen}. A detailed \geantIII-based detector and
 trigger simulation is used to produce the detector response, which 
 is processed using the same reconstruction algorithms as data. In
 addition to a \btodpi sample, we also use samples of three other decay
 modes; $B^0 \to D^+\pi^-$ ($D^+ \to K^- \pi^+\pi^+)$ , $B_s \to \phi\phi$ and $B_s \to K^+K^-$, where
 the offline selection criteria applied are broadly similar to that of
 the \btodpi candidates. These distinct samples, with differing topologies, allow for further
 crosschecks of the basis of the method to correct the selection biases. The calculation of the efficiency
 function is easily extended to include four track decays using the
 same principle of scanning through all possible proper decay times as
 described in Sec.~\ref{sec:removeBiasForSignal}.

 As these samples contain only signal events, we use the PDF described
 in \equationref{eq:withSVTProb2} to simultaneously extract the
 lifetime and the L2 single track finding efficiency. The fitted
 lifetimes, along with the input truth lifetimes and size of each
 sample are given in Table~\ref{tab:realMCresults}.
\begin{table*}[htbp]

\caption[Full MC fits]{The fit results on full detector simulated $B$ decay samples. The table also gives the true input lifetime and the size of the sample after selection cuts had been applied.} \label{tab:realMCresults}
\begin{center}
\begin{tabular}{lccc}
\hline \hline Decay &Sample size & Input lifetime & Measured lifetime\\ \hline 
 \btodpi  & 75000 & 496 $\mu$m   & 493.3 $\pm$ 3.2 $\mu$m  \\ 
$B^0 \to D^+ \pi^-$ & 71000  & 464 $\mu$m & 467.8 $\pm$ 2.8 $\mu$m  \\ 
$B_s \to \phi \phi$& 35000  & 438 $\mu$m & 443 $\pm$5 $\mu$m      \\
$B_s \to K^+ K^-$& 75000  & 438 $\mu$m  & 441.5$\pm$ 2.9 $\mu$m   \\ \hline \hline
\end{tabular}
\end{center}

\end{table*}
The fitted lifetime is consistent with the
input lifetime for each Monte Carlo sample. These results indicate that the method of calculating
the event efficiency can be used to correct the selection biases.

\subsection{The Fast Simulation}
\label{sec:fastSimulation}

 In addition to the full CDF II detector simulation we use a custom fast
 simulation which is several orders of magnitude faster than the
 detailed simulation. It allows production of many thousands of
 independent samples, each approximately the size of the data yield
 ($24,000$ signal events), that are used for the extensive validation
 and studies of systematic uncertainty. The fast simulation is used for validating
 the technique with simulated signal and background events, and for
 evaluating systematic uncertainties. Neither the fast simulation nor
 the full simulation described earlier is used to determine or
 constrain any of the parameters that enter the likelihood fit to data
 from which we extract the \bplus\ lifetime.
 Below, we describe the fast simulation with its default
 settings. These form the basis of the validation studies presented
 later.  How the default behavior is altered to estimate systematic
 uncertainties is discussed in \secref{sec:Systematics}.

 In order to reproduce the data as well as possible with a relatively
 simple simulation, we generate many of the kinematic variables in
 each event based on distributions observed in data, in particular
 when generating background. The most important ones are summarized in
 \tabref{tab:fastSim}.
 \begin{table*}
 \caption{Kinematic parameters of the fast simulation and the
 parent distribution used for generation. Details are given in the text.\label{tab:fastSim}}

 \begin{tabular}{cllc}
 \hline\hline

 \multicolumn{3}{c}{Randomly generated parameter} & Parent distribution \\\hline
 $t$ & reconstructed decay time of $B^-$ & (signal) & $\frac{1}{\tau_B}
 e^{-t/\tau_B} \otimes \gauss{t}{\sigma_t}$\\
 $t_D$ & reconstructed decay time of $D^0$ & (signal) & $\frac{1}{\tau_D}
 e^{-t_D/\tau_D} \otimes \gauss{t_D}{\sigma_t}$ \\
 $|\vect{P}|$ & magnitude of $B^-$ momentum & (signal)  &
 background-subtracted data \\
 $m_B$ & $B^-$ mass &(signal) & PDF given in \equationref{e:masspdf},
 fitted to data\\ 
 &&&\\ 
 $t$ & reconstructed decay time of $B^-$ & (bkg) & PDF given in \equationref{e:bkglife}, fitted to data \\
 $t_D$ & reconstructed decay time of $D^0$ & (bkg) & sideband data \\
 $|\vect{P}|$ & magnitude of $B^-$ momentum & (bkg)  &
 sideband data \\
 $m_B$ & $B^-$ mass &(bkg)& PDF given in \equationref{e:massbkgpdf}, fitted
 to data\\ 
 &&&\\
 $\phi$ & azimuth angle of $B^-$ momentum & & uniform \\ 
 $\eta$ & pseudorapidity of $B^-$  & & uniform with $|\eta|<1.5$ \\
 $\Delta d_0^{L2}$ & \multicolumn{2}{l}{ \dzSVT$-$\dzOff} & Gaussian, then round
 \\ &&&   \dzSVT\ to nearest \un{10}{\mu m}.
 \\\hline\hline
 \end{tabular}
 \end{table*} 
 For every event $i$ we generate the $B^-$ proper decay time, $\tm$,
 the reconstructed mass, $\mass$, the measured momentum, $\vect{P_i}$,
 and the $D^0$ meson proper decay time.  The $B^-$ mass is generated
 from the PDF described in ~\equationref{e:masspdf} using the best fit
 parameters from the mass fit to the data sample. For signal events,
 the $B^-$ and $D^0$ proper decay times are generated as exponentials
 using the 2008 world average values of the lifetimes, which
 are 1.637~ps and 0.41~ps for the $B^-$ and $D^0$ mesons,
 respectively~\cite{PDG}. The generated proper decay times are smeared
 by a Gaussian of width 0.087~ps to simulate the detector
 resolution. The generation of the reconstructed $B^-$ proper decay
 time in background events is based on the PDF described in
 \secref{p:time}. Its parameters are determined from data, by fitting
 the lifetime distribution of the events in the upper mass sideband.
 The background $D^0$ proper decay time is taken from the $D^0$ decay
 time distribution observed in the upper mass sideband. The direction of the
 $B^-$ momentum is generated uniformly in $\phi$ and $\eta$.  As
 transverse quantities are used to determine the measured proper decay
 time in data, it is important to match the $p_T$ distribution in the simulation, to that observed in data. The magnitude of the $B$ momentum is generated such that, after the
 selection criteria are applied, the distribution of $p_T$ of the
 remaining simulated signal events matches the $p_T$ distribution
 observed in the background subtracted signal region. Similarly, we
 generate the magnitude of the momentum for background events so that
 after selection there is agreement between the $p_T$ of simulated
 events and the upper sideband in data.

 We calculate the remaining kinematic variables as follows. In the
 rest frame of the $B^-$ particle, the magnitudes of the reconstructed
 $D^0$ and $\pi_B$ momenta are defined by the generated mass of the
 $B^-$ meson and the world average values for the $D^0$ and $\pi$
 masses~\cite{PDG}. The reconstructed $D^0$ mass is kept fixed because,
 in data, the mass-constrained vertex forces the reconstructed $D^0$
 mass to the world average value.  We pick a direction for the $\pi_B$
 momentum isotropic in the $B^-$ rest frame; the $D^0$ momentum is in
 the opposite direction. These momenta are then transformed into the
 laboratory frame to calculate the simulated $D^0$ and $\pi_B$
 momenta. The equivalent procedure is carried out to calculate the
 $\pi_D$ and $K_D$ momenta in the laboratory frame. The $B^-$ and
 $D^0$ decay vertex positions are calculated from the generated proper decay
 time and momentum; knowledge of these allows for track impact
 parameter calculation. These impact parameters are defined to be the
 offline impact parameters.

 We simulate the SVT with a single track finding efficiency of
 $\effsg=65\%$ for signal events and
 $\effbkg=55\%$ for background events. The efficiency
 is different for signal and background because, in general, we find in
 our data that background tracks have fewer hits in the silicon layers,
 and hence a lower track finding efficiency. The values for the track
 finding efficiency we use for the simulation are approximately those
 found in data for tracks with \un{|\dzOff| <1000}{\mu m}, obtained from the
 simultaneous proper decay time, mass, and efficiency fit (the fit
 results for all parameters can be found in
 \appref{sec:fittab}. Simulation tracks with \un{|\dzOff| > 1000}{\mu m} are not used in the trigger decision, and are treated in the fit as
 not found by the SVT, so there is no need to model the behavior of
 the SVT efficiency for tracks with \un{|\dzOff| > 1000}{\mu m}.
 For those tracks that are found, the SVT-measured impact parameter,
 $\dzSVT$ is obtained by adding a Gaussian-distributed random number
 to the \dzOff. The Gaussian is centered at 0 and has a width
 of \un{35}{\mu m}, which is consistent with the width
 observed in data (\figref{f:widths}).  The result is then rounded
 to the nearest \un{10}{\mu m}, as in the real SVT.  The difference
 between the L3 impact parameter $\dzLt$ and $\dzOff$ is not simulated. Although the mean $\dzSVT$ in data is shifted from zero, further tests, detailed below, confirm that the central value of the $\dzSVT$ distribution does not affect the results. Therefore, these differences between the fast simulation and the data will have a negligible effect on the interpretation of the results.

\begin{figure}[htbp]
\begin{center}
\subfigure[]
{\includegraphics[scale=0.35]{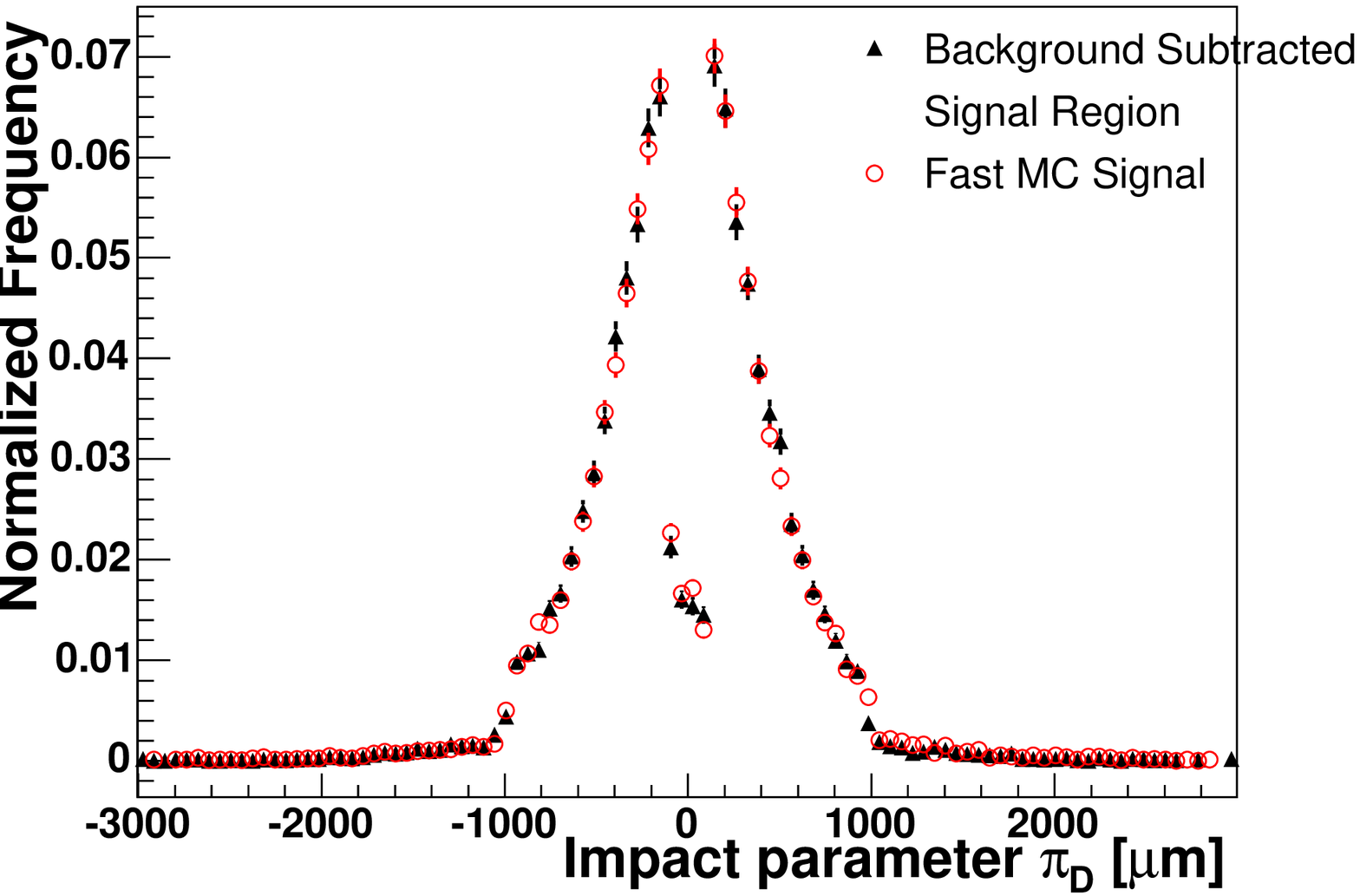}
\label{fig:IPsgcom}}
\qquad
\subfigure[]
{\includegraphics[scale=0.35]{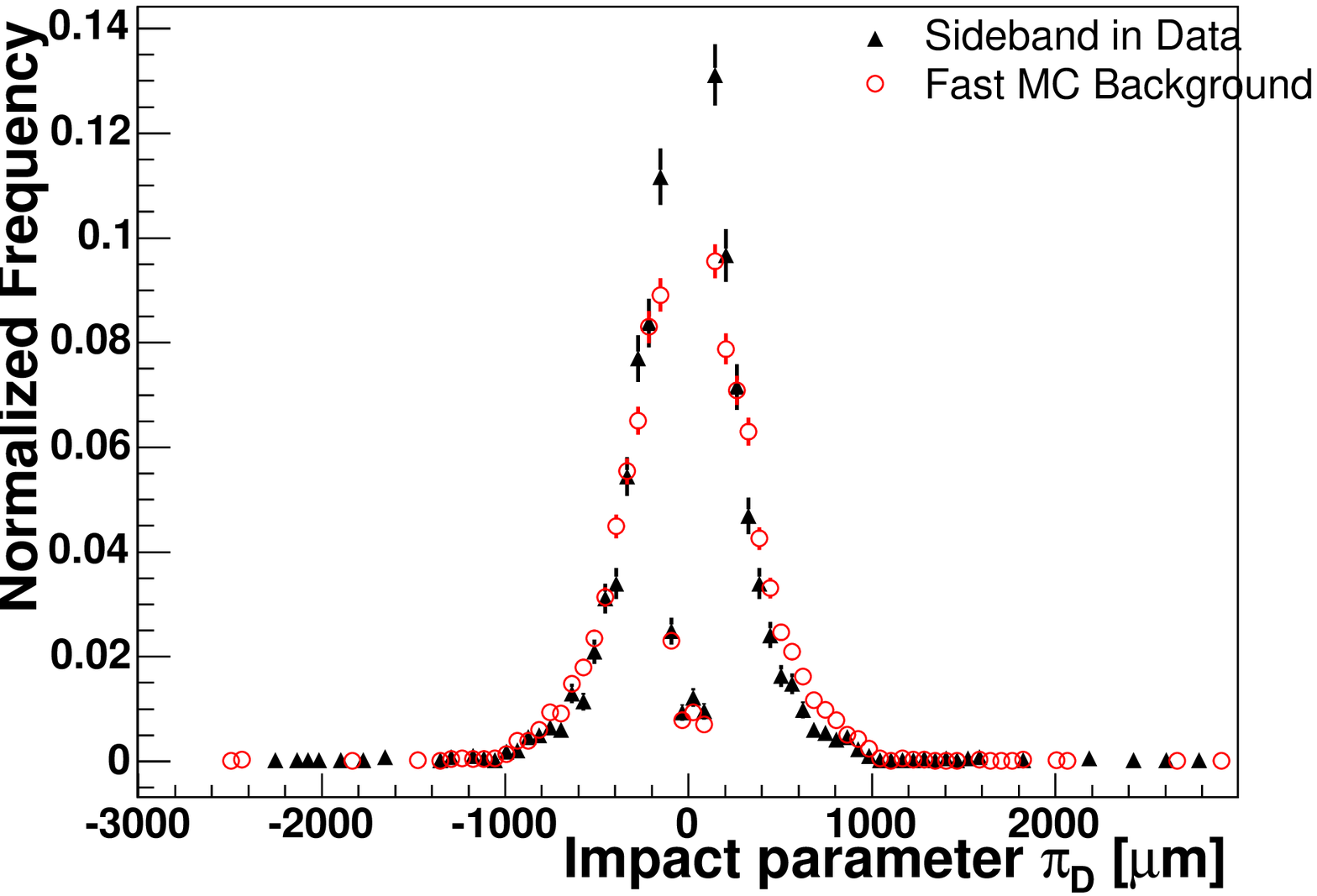}
\label{fig:IPbgcom}}
\caption[]{A comparison of the impact parameter distribution of the $\pi_D$ track in data (triangle points) and fast simulation (circular points). The comparison between generated signal events and the background subtracted signal region in data is shown in (a) while (b) shows the comparison between generated background events and the upper sideband in data. All distributions are normalized to one event. \label{fig:IPcomp}}
\end{center}
\end{figure}

 After all kinematic quantities have been obtained in the way
 described above, the selection criteria are applied to replicate the
 biases observed in data. All decay products are required to lie in
 the fiducial volume of the CDF II detector. The three two-track trigger configurations summarized in Table~\ref{secDetector:tab:trigcut} represent three different sets of selection criteria. Events
 are generated with each set of cuts separately and then combined in
 the fractions observed in the data. In data we observe very few
 events with tracks that have $|\eta|>1.5$. Therefore events that have
 simulated tracks with $|\eta|>1.5$ are removed from the sample. For
 background, prior to applying selection cuts, we further reject events so that the $p_T$ spectrum of
 the candidate $\pi_B$ after the cuts are applied matches that observed in the data upper
 sideband. This further rejection for background events effectively
 changes all kinematic distributions observed after the selection
 criteria are applied and forces the simulated background to have the
 characteristics of background observed in data.  Overall there is
 broad agreement between the distributions of impact parameters,
 momenta, $\Delta\phi$ of track pairs, and $\eta$ in the simulated
 and real data. As impact parameters are particularly important in
 this analysis, we compare the $\pi_D$ impact parameter distribution
 from the fast simulation and in real data in
 Fig.~\ref{fig:IPcomp}. Given the simple nature of the fast
 simulation, the agreement with data is remarkably good, although of
 course not perfect. Since the simulation is not used to determine any
 parameters in the final fit to data, but only to test the robustness
 of the method and to estimate systematic uncertainties, we do not
 rely on a perfect match between the simulation and the data, and the
 agreement we observe is sufficient.

\subsection{Validation of the method on signal events}
\label{sec:sigval}
 
 We use the custom fast simulation for high-statistics tests of the
 signal PDF given in Eq.~(\ref{eq:withSVTProb2}). We generate 1000
 samples of $24,000$ signal events each, similar to the yield observed in
 data. The proper decay time distribution for each sample has been
 sculpted by the same decay time dependent selection cuts as in real data,
 applied to the simulated data as described in the previous
 section. We maximize a likelihood function for signal events, constructed from the PDF in \equationref{eq:withSVTProb2}, to
 extract a best fit lifetime for each sample. Fitting the resulting
 pull distribution with a Gaussian, we find a mean $\mu = -0.026 \pm
 0.034$ and a standard deviation $\sigma = 1.027 \pm 0.024$. This
 demonstrates that \acc\ is correctly calculated, and that the
 likelihood formed from the PDF in \equationref{eq:withSVTProb2} can
 correct for the selection biases. It also shows that assigning the
 value of $\delDz$ to tracks that did not have an SVT match, from the
 distribution of $\delDz$ of tracks that did have an SVT match, does not cause any
 bias. In addition to this single test to validate the method itself
 we performed further tests to cross check our assumptions described below.

 There are some differences between the value of the single track
 finding efficiency applied in the fast simulation and the efficiency observed
 in data. To test that the results were not sensitive to the default
 values of the efficiency chosen for the fast simulation, we varied
 the input efficiencies around the default values and saw no bias due
 to the value of input efficiency or due to the difference between signal and
 background efficiencies. The fitted efficiency was always consistent with the input value.
 In Sec.~\ref{sec:trigselec} we noted that there have been three
 changes to \effSVT\ over the course of the period of data taking used
 for this analysis. 
To determine whether it is sufficient to parametrize the SVT track finding efficiency with a single value (representing the average \effSVT\ over these three data taking periods), we generated samples containing events simulated using three different values of \effSVT\ in the proportions observed in data.
These samples were fit using only one average \effSVT\
 parameter, which was allowed to float in the fit. The resulting pull
 distribution has a mean consistent with 0~$\mu$m, however the width is 1.19$\pm$0.03. This can be understood as
 follows: Each \acc\ is a measure of the statistical power of each
 event~\cite{Rademacker:2005ay}. By using an average $\effSVT$, the
 statistical power of each event has been incorrectly assumed in the
 fit leading to an incorrect estimate of the statistical
 uncertainty. If instead, we allow for three floating efficiency
 parameters where each parameter is only sensitive to the events in
 one of the data taking periods, the resulting pull distribution once again
 has unit width. Therefore, in the fit to data, we use three parameters
 to describe $\effSVT$, each floating in the fit, one for each data
 taking period.

 In the default simulation the $(\Delta d_0)_{L2}$ distribution is generated with a
 Gaussian distribution with mean $\mu=0$~${\mu m}$ and width $\sigma=\un{35}{\mu m}$.
 As the fit method takes all its information about $(\Delta d_0)_{L2}$ from
 data, and makes no assumptions about the shape of the $(\Delta d_0)_{L2}$
 distribution, we expect it to perform equally well for any $(\Delta d_0)_{L2}$
 distribution, including asymmetric and biased distributions. We test
 this by generating data with two alternative models for $(\Delta d_0)_{L2}$: For the first model we use a
 biased impact parameter resolution function described by a Gaussian with mean $\mu =
 \un{35}{\mu m}$ and width $\sigma = \un{35}{\mu m}$. To truly
 stress-test the sensitivity of the method to the $(\Delta d_0)_{L2}$
 distribution, the second alternative model is a, somewhat unrealistic, biased and asymmetric resolution function
 described by an exponential decay distribution with mean
 $\un{35}{\mu m}$, so that all \dzSVT\ are larger than the \dzOff .
 For both models we perform pull studies with the same sample size as observed in data and observe no bias in the fitted lifetime.
 This confirms that the fit method is robust with respect to the shape
 and mean of the $(\Delta d_0)_{L2}$ resolution function, and that the observed
 shift from zero in data of the mean of the distribution in \figref{f:means} does
 not affect the fit result.

 Other assumptions, including the dependence of the SVT efficiency on
 impact parameter, $p_T$, $\eta$, and the effect of small differences
 in the $(\Delta d_0)_{L2}$ resolution depending on impact parameter, are
 discussed as sources of systematic uncertainties in
 Sec.~\ref{sec:Systematics}.

%-------------------------------------------------------------

\section{The Combined PDF for Signal and Background Events}
\label{sec:SgBgPDF}
In this section we derive the PDF for a sample containing signal and
background events. We remind the reader that we use four measured
observables in the fit; the measured proper decay time, $t_i$, the
efficiency function, $\acc$, the mass, $\mass$, and the
track-configuration observed, $\trk$.  An unbinned maximum likelihood
fit is used to determine the lifetime of the $B$ meson and other
parameters. Candidates in the data sample have passed the selection
criteria, \trig, which means that we must consider the conditional
probability that a candidate has a particular \tm, \mass, \acc, and \trk,
given that the selection criteria have been satisfied. There are only
two classes of candidates in the data sample: signal and background;
therefore, the likelihood function is defined as
\begin{equation}
\begin{split}
\label{eq:fullPDF:sb}
 \mathcal{L}
 = \prod_i 
    \big[ &
              \mathcal{P}\left(s, \tm, \mass, \trk, \acc;\tau | \trig \right)
       \\ & + \mathcal{P}\left(b, \tm,  \mass, \trk, \accb | \trig \right)
    \big],
\end{split}
\end{equation}
where the first term represents the likelihood for signal candidates and
the second term is the likelihood for background candidates.  For
readability, the dependence on other fit parameters, such as those
related to the parametrization of the mass distribution, is suppressed
and only the dependence on the fit parameter $\tau$ is explicitly
written.

The PDF for signal candidates can be factorized into the following form,
\begin{equation}
\begin{split}
\label{eq:fullpdf}
  \mathcal{P}&(s, \tm, \mass, \trk, \acc;\tau | \trig )=
\mbox{}
\\ &   \mathcal{P}(\tm;\tau|\trig, \acc, s)
\\ & \times \mathcal{P}(\trk |\trig, \acc, \tm, s) 
\\ & \times \mathcal{P}(\acc |\trig)\times\mathcal{P}(\mass|\trig,s)
\\ & \times \mathcal{P}(s|\acc,\trig) 
\end{split}
\end{equation}
 where a detailed derivation of this factorization is given in
 \appref{sec:Derivation}. There is also an entirely analogous factorization for background candidates. The combined factor $\mathcal{P}(\tm;\tau|\trig, \acc,
 s)\mathcal{P}(\trk |\trig, \acc, \tm, s)$ describes the proper decay
 time distribution and includes the track configuration information
 which determines $\effSVT$. 
 Note that $\mathcal{P}(\acc|\trig)$ and similar expressions refer to
 the probability to find a given efficiency \emph{function}~\acc. It
 does not refer to the function as evaluated for a given $t$ or
 $\effSVT$, but to the function as a whole. $\mathcal{P}(\acc|\trig)$ therefore
 does not depend on the value of $t_i$ or $\effSVT$.
 The factor $\mathcal{P}(\acc|\trig)$ is independent of $\tau$ and
 whether a candidate is signal or background. Hence, it can be ignored in
 the likelihood. The factors $\mathcal{P}(\mass|\trig,s)$ and
 $\mathcal{P}(\mass|\trig,b)$ (from the background part of the PDF)
 describe the mass distribution and are described
 earlier, in Sec.~\ref{sec:sampcomp}. The final factor
 $\mathcal{P}(s|\acc,\trig)$ is the probability that a candidate is
 signal given its efficiency function. Each factor of the likelihood
 is normalised to one candidate.
\subsection{The parametrization of the background proper decay time PDF}
\label{p:time}
 This section considers the proper decay time term in the PDF in
 \equationref{eq:fullpdf}, the analogous term for the background
 candidates, and describes the parametrizations of the PDFs used for the fit. For
 the signal component a physics model is used, for the background
 contribution it is sufficient to provide an empirical description of
 the data. The first two factors on the right hand side of
 \equationref{eq:fullpdf} are identical to the left hand side of
 \equationref{eq:withSVTProb2}, and this is the PDF used to fit the
 proper decay time and single track finding efficiency for signal
 candidates. Three different values of \effsg\ are fit, one for each
 time period as described in Sec.~\ref{sec:validation}.

 For background candidates 
\begin{equation}
\begin{split}
  \mathcal{P}(\tm;\tau| & \trig, \acc, b)\mathcal{P}(\trk |\trig,\acc, \tm, b)
\\ & = \frac{\effbg^n(1-\effbg)^{n-r} y(\tm)}{ \sum\limits_{\substack{k=all\\intervals}}
    H_k(\effbg) \int\limits_{t_{\mathrm{min\; k}}}^{t_{\mathrm{max\; k}}} y(t)dt},
\end{split}
\end{equation}
where, similarly to signal, there are three values of \effbg\ to fit.
The function $y(t)$ can be determined empirically from the data. Simple forms of $y(t)$, such as a sum of exponentials convoluted by a Gaussian, were found to provide an unsatisfactory description of the data.
Therefore, the function $y(t)$ is empirically determined using an interpolation of exponentials given by

\begin{equation}
	y(t) = e^ 
			{a_j + \left ( \frac{
						a_{j+1} - a_j }{
						t_{j+1} - t_j } \right )
			   (t - t_j)
			}    \; \mathrm{for}\; t_j \leq t \leq t_{j+1}.
\label{e:bkglife}
\end{equation}
We use ten fit points ($t_j$), which are spaced more closely at low $t$ where the proper decay time distribution of background candidates is concentrated. The values of the corresponding $a_j$ are determined alongside the other fit parameters in the unbinned maximum likelihood fit. This parametrization was tested on data from the upper sideband to ensure that it is a good model for the data. The tests on the upper sideband were only used to distinguish the performance of different parametrizations. No fit parameters are fixed from this test.

\subsection{The complication in combining the signal PDF and the background PDF when using a candidate-by-candidate efficiency function}
\label{p:fishwhy}

Combining the signal and the background PDF while using a candidate-by-candidate
efficiency function introduces a significant complication into the
analysis. The rest of this section describes this problem, and its
solution, in detail. As discussed in~\cite{Punzi:2004wh}, when a candidate-by-candidate quantity
enters a fit with a signal and background component, the PDF for this quantity needs to be included in the fit. In
our PDF this effect is taken into
account by a term that describes the candidate-by-candidate signal probability
dependent on the \acc. So, instead of an overall signal
fraction $P(s)$, there is a signal weighting for each candidate which
depends on the efficiency function. This is described by the factor
$\mathcal{P}(s|\acc, \trig)$, and the corresponding term for
background is simply
$\mathcal{P}(b|\acc,\trig)=1-\mathcal{P}(s|\acc,\trig)$.  Alternative
ways of factorizing the PDF would lead to different ways to take this
effect into account, but, regardless of
the choice of factorization, the underlying need to include a
PDF for the efficiency function remains.

\begin{figure}
\begin{center}
\resizebox{1.0\columnwidth}{0.2\textheight}{\includegraphics[clip]{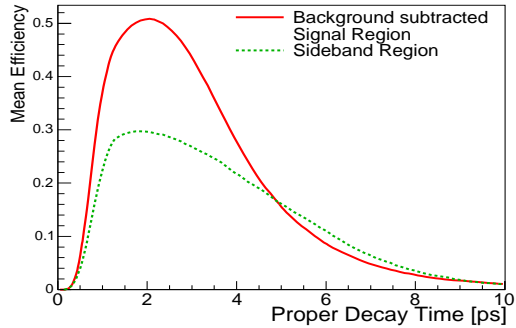}}
\end{center}
\caption[Background proper decay time fit projection]{The mean $E(t,\effSVT)$ function for signal and background candidate. Signal (solid line) and background (dashed line) candidates have different $\overline{\acc}$.}
\label{fig:meanacc}

\end{figure}

The factor $\mathcal{P}(s|\acc,\trig)$ can be simplified to an overall signal fraction, $\mathcal{P}(s)$, only in the case where the efficiency function distributions
are the same for signal and background. Figure~\ref{fig:meanacc} shows
the mean efficiency function, $\overline{\acc}$ for candidates in the upper sideband
(background) and the background subtracted signal region. The mean is determined simply by summing all efficiency functions in a sample and dividing by the number of candidates. The two $\overline{\acc}$ are clearly different which shows that the distribution
of efficiency functions in signal and background must be different. We can
estimate the bias on the lifetime measurement we would get if we were to ignore the
differences in the efficiency function, by simplifying
$\mathcal{P}(s|\acc,\trig)$ to $\mathcal{P}(s)$. Using the custom fast
simulation described in \secref{sec:fastSimulation}, we find a bias of
approximately $-$0.018$\pm$0.001~ps. Any advantage gained in precision by using a
simulation independent method would be negated by a bias of this
size. Therefore, a successful simulation-independent method for
correcting a trigger bias must include a proper description of
the term $\mathcal{P}(s|\acc,\trig)$.

\subsection{Calculating the term  $\mathcal{P}(s|\acc,\trig)$}
\label{s:Psgivenacc}
\subsubsection{Overview}
To correctly represent the PDF in the fit we require a
parametrization of the signal fraction that is dependent on the
candidate-by-candidate efficiency function, \acc. However, it is difficult to parametrize a distribution of functions, and that is what is required to derive a signal probability as a function of each individual efficiency function. The problem
is simplified if we represent \acc\ by a
number, $x_i$, as it is considerably easier to parametrize the
distribution of the scalar variable $x$ rather than a distribution of
functions, i.e., we aim to find a variable $x$ such that we can replace
$\mathcal{P}(s|\acc,\trig)$ by assuming $\mathcal{P}(s|x,\trig) \approx
\mathcal{P}(s|\acc,\trig)$. For this approach to succeed, $x$ must be
chosen in such a way that the loss of information regarding the signal
probability contained within $\acc$ is minimized as we transform from $\acc$ to
$x$.  Note that the transformation of \acc\ to $x$ is only used for
determining the signal probability of each candidate. The proper decay time
probabilities are unchanged and continue to use \acc, as the
trigger bias cannot be corrected without the full description.
To summarize, the parametrization of the term $\mathcal{P}(s|\acc,\trig)$
involves two steps:
\begin{itemize}
\item Transforming the efficiency function $\acc$ into a
  representative number $x$.
\item Describing the signal fraction as a function of $x$, $P(s|x, \trig)$
  with a suitable function whose parameters will be determined in the fit.
\end{itemize}
These are discussed below.

\subsubsection{Representing the Efficiency function by a Scalar}
\label{s:scalar}
\label{p:accvsx}
In order to translate the efficiency function \acc\ into a scalar
variable, we make use of the Fisher Linear Discriminants
method~\cite{Fisher1}.  This method transforms a vector of variables
into a single scalar variable. We represent each
\acc\ as a vector that contains all the relevant
information about \acc\ and then use the Fisher discriminant method to
translate this vector into a number, the Fisher scalar, $x$. Note that we
do not use the Fisher discriminant method to select candidates.
The scalar resulting from the Fisher discriminant method is optimized
for distinguishing signal from background, and therefore fulfils the
requirement of minimizing the loss of information about the signal
probability as we translate \acc\ to $x_i$, so that $P(s|x_i, \trig)
\approx P(s|\acc, \trig)$ to a very good approximation. How good this
approximation is, is quantified below.
Here, we summarize the method rather briefly. Further details can be
found in \appref{sec:FisherAppendix}.

\subsubsection{Finding the Fisher discriminant in a simulation independent way}
The Fisher scalar variable, $x_i$, is given by $x_i = \vect{w}\cdot \vect{v}_i$ where $\vect{v}_i$ represents one candidate's efficiency function and $\vect{w}$ is another fixed vector. The Fisher Linear
Discriminant method provides a way to determine a vector $\vect{w}$, such that it
maximizes the separation of signal and background candidates in the
variable $x$. The transformation of the information contained in \acc\ to $\vect{v}_i$ is described in detail in \appref{sec:FisherAppendix}. The transformation does not require the values of \effsg\ or \effbg\ and hence the transformation can be done before the fit determines the values for the efficiencies.

In typical uses of the Fisher Linear Discriminant method, the
calculation of $\vect{w}$ requires not only the knowledge of all the
$\vect{v}_i$, but also knowledge of $\vect{v}_s$ and $\vect{v}_b$, which are the
mean $\vect{v}_i$ for signal and background candidates,
respectively. Traditionally, $\vect{v}_s$ and $\vect{v}_b$ are
determined from independent training samples, such as detailed Monte
Carlo data. Since this analysis uses no input from simulation
we use the data itself to calculate $\vect{v}_s$ and
$\vect{v}_b$. For this measurement, we use candidates in the upper sideband to
determine $\vect{v}_b$. We perform a background subtraction on candidates
with $5.25<m_B<5.32$~GeV/$c^2$ to determine $\vect{v}_s$. Further information regarding the determination of $\vect{v}_b$ and $\vect{v}_s$ is given in \appref{sec:FisherAppendix}.

\subsubsection{Testing the assumption that $P(s|x_i, T) \approx P(s |
  \acc, T)$}
\label{sec:FisherTestPofAccIsPofX}
Before proceeding further, it is important to test the assumption that
the Fisher scalar variable $x_i$ is representative of $\acc$. We use a
custom fast simulation and fit the lifetime of the 1000 independent
samples of signal and background candidates, using the Fisher scalar $x_i$
to determine a signal probability per candidate. It is desirable to quantify how the assumption that $P(s | \acc, T) \approx P(s|x_i, T)$
affects the fit result, in a way that is independent of any particular
parametrization of $P(s| x_i, T)$. (The particular choice of parametrization is discussed separately and is described in Sec.~\ref{p:fish}.) To do this we make use of the truth information available from the simulated data. As shown by the data points in Fig.~\ref{fig:fitfish}, $\mathcal{P}(s|x_i,\trig)$
can be calculated by finely dividing the sample into 100 bins in $x$, and simply counting the number of signal and background candidates in
any particular bin of the variable $x$. So, for each $x_i$, we
determine $\mathcal{P}(s|x_i,\trig)$ by reading its value off a
histogram generated from the truth information. We find that the mean lifetime shift in those 1000 fits is
only 0.0013~ps, which is significantly smaller than $-$0.018~ps found
when the distribution of efficiency functions is ignored. This
demonstrates that the variable $x_i$ is a satisfactory substitute for
\acc\ for the purposes of calculating the probability that a candidate is
signal given its efficiency function, and that $P(s|\acc,\trig)\approx P(s|x_i,\trig)$ is a reasonable assumption. This mean shift of 0.0013 ps is small in comparison to the statistical uncertainty from the data sample size and is
taken as the systematic uncertainty due to assuming the scalar variable is entirely equivalent to using the full efficiency function. This method of calculating $\mathcal{P}(s|x_i,\trig)$ is only used in this set of test fits. For other tests, (and for the final data fit), no truth information is used, and the parameterization of $\mathcal{P}(s|x_i,\trig)$ described in \secref{p:fish} is used.

\subsubsection{Parametrizing the signal fraction as a function of the
  Fisher scalar variable}
\label{p:fish}

In order to apply this method to real data in a simulation-independent
way, we need to find a function that parametrizes $P(s|x_i,\trig)$, and
whose parameters can then be determined in the fit to data.
We use Lagrange interpolating polynomials as they provide a very
general parametrization that makes minimal assumptions about the
shape of the distribution to be fitted. This parametrization has as
its parameters the signal fractions $p_j$ at certain discrete values
of the Fisher discriminant $x_j$, so $P(s|x_j,\trig) = p_j$. The value
of $P(s|x, \trig)$, for general $x$, is calculated using a smooth
interpolation between those points. The $p_j$ are determined in the
fit.

Our default choice for the $x_j$ is the following: We divide the
$x$~axis into $N=15$ equal bins.  As the number of candidates at
the edges of the distribution is small, we merge the first two
bins, and also as the last two bins. We place our $x_j$ at
the center of each of the resulting bins. This results in 13 fit
parameters, $p_j$, representing the signal fractions at the 13 $x_j$. We
tested the robustness of this choice by trying out different numbers
of bins $N$, and found that there is negligible difference in
performance for any value of $N$ from 10 to 20.

\begin{figure}[htbp]
\begin{center}

\includegraphics[width=\columnwidth]{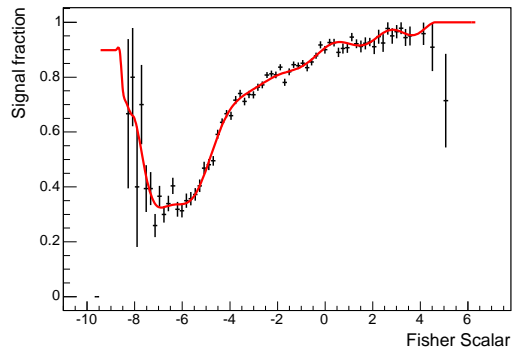}
\end{center}
\caption[Signal fraction as a function of Fisher scalar]{The data
  points show the signal fraction as a function of Fisher scalar for a
  sample of simulated data. The line shows the projection of the
  Lagrange interpolating polynomial determined by the simultaneous fit
  to proper decay time, signal fraction and other parameters. \label{fig:fitfish}}

\end{figure}

This parametrization is tested using the fast
simulation. Figure~\ref{fig:fitfish} shows the projection of the
fitted Lagrange interpolating polynomial, $f(x)$, where the truth information
has been superimposed for one sample of simulated data. In contrast to
the test in \secref{sec:FisherTestPofAccIsPofX} where we tested the assumption
$P(s|x_i, \trig)\approx P(s|\acc,\trig)$, the fit, here, is performed in
the same way as in our final fit to real data: at no point is truth
information or any external simulation input used in the fit, and the
$p_j$ parameters of $P(s|x_i,\trig)$ are determined in the fit at the
same time as all other fit parameters, such as the lifetime or
$\effSVT$. The projection of $P(s|x,\trig)$ obtained in this fit
matches closely the histogram obtained from truth information,
giving us confidence that this parametrization provides a good
description.  We tested this parametrization using 1000 simulated
samples and observed a mean residual of 0.0013 ps. The lifetime pull
distribution is described by a Gaussian with mean 0.039$\pm$0.036 and
width 1.097$\pm$0.029. This demonstrates no further shift in the mean residual position relative
to the small shift, resulting from the assumption $P(s|x_i,
\trig)\approx P(s|\acc,\trig)$, observed in \secref{sec:FisherTestPofAccIsPofX}. As the parametrization works as well as the truth information any systematic uncertainty due to the parametrization of $P(s|x_i,\trig)$ is negligible.

\subsubsection{Summary: The full signal \&\ background PDF with the factor  $\mathcal{P}(s|\acc, \trig)$ }
\label{p:fishsummary}
In summary, we find that the PDF in \equationref{eq:fullpdf}, with the
factor $P(s|\acc,\trig)$ parametrized as described in this section,
successfully corrects for the selection bias in data samples where
both a signal and a background component is present. The 0.0013~ps
residual is taken as a systematic uncertainty due to the method of
describing the term $\mathcal{P}(s|\acc, \trig)$ by the $x_i$ variable. The width of the
pull indicates that the method underestimates the statistical
uncertainty by 10$\pm$3 $\%$. To be conservative we increase the
statistical uncertainty of the fit to data accordingly.

%------------------------------------------------------------

%------------------------------------------------------------
\section{Fit Results}
\label{sec:FitResults}

This section describes the fit to data selected by applying the
 selection criteria listed in \secref{sec:btodpi}. An initial
 mass fit is performed, as described in \secref{sec:sampcomp}, with
 seven free parameters. The best fit results are given in \appref{sec:fittab}. The results of the mass fit are used to
 perform the background subtraction required to calculate the $\vect{v_s}$, which is needed for the Fisher Discriminant Analysis. 

 The lifetime is determined in a second fit. The likelihood function used in this unbinned maximum likelihood fit is given by
 \equationref{eq:fullPDF:sb} and \equationref{eq:fullpdf} and is
\begin{widetext}
\begin{equation}
\begin{split}
\log \mathcal{L} &=\sum_i \log \Biggl[ \frac{ f(x)|_{x=x_i}\cdot \effSVT^r(1-\effSVT)^{(n-r)}\frac{1}{\tau}e^{\frac{-t_i}{\tau} + \frac{1}{2} \frac{\sigma_t^2}{\tau^2}}\Freq\left( \frac{t_i}{\sigma_t} - \frac{\sigma_t}{\tau}\right)
   }{\intervSum\mathrm H_{k_i}(\effSVT)\left[
           - e^{\frac{-t}{\tau} + \frac{1}{2} \frac{\sigma_t^2}{\tau^2}}
           \Freq\left( \frac{t}{\sigma_t} - \frac{\sigma_t}{\tau}\right)
           + \Freq\left(\frac{t}{\sigma_t}\right)
    \right]_{t=t_{\mathrm{min\; {k_i}}}}^{t=t_{\mathrm{max\; {k_i}}}}} \\
& \quad \times  \left( \frac{f_1\cdot  e^{ - \frac{\left( {\mass - m_1 } \right)^2 }{2\sigma_1^2}}}{\sigma_1 \sqrt{2\pi}}  +\frac{(1-f_1)\cdot e^{ - \frac{\left( {\mass - m_2 } \right)^2 }{2\sigma_2 ^2 }} }{{\sigma_2 \sqrt {2\pi } }}\right) \\
& + \frac{\left(1-f(x)|_{x=x_i}\right)\cdot\effbg^n(1-\effbg)^{n-r} y(\tm)}{ \sum\limits_{\substack{k=all\\intervals}}
    H_k(\effbg) \int\limits_{t_{\mathrm{min\; k}}}^{t_{\mathrm{max\; k}}} y(t)dt}\cdot\frac{1-\alpha \mass}{\left[ m_{\mathrm{high}}-m_{\mathrm{low}} -\frac{\alpha}{2}\left(m_{\mathrm{high}}^2-m_{\mathrm{low}}^2\right)\right]} \Biggr], 
\end{split}
\end{equation}
\end{widetext}
where $y(t)$ is defined in \equationref{e:bkglife} and $f(x)$ is described in \secref{p:fish}. The parameters that determine the mass
 shapes for signal and background are fixed at the values
 determined in the initial mass fit. However, the signal fraction is not taken from the mass fit because this is now redefined in terms on the
 Fisher scalar variable. In total, there are 30 free parameters in the
 lifetime fit. These are the following: one for the signal lifetime, ten to
 describe the background proper decay time distribution as described
 in Sec.~\ref{p:time}, 13 parameters to determine the signal
 fraction as a function of the Fisher scalar, $f(x)$, defined in
 Sec.~\ref{p:fish} and six parameters to describe the single track
 finding efficiency as described in Sec.~\ref{p:time}.

\begin{figure*}
\begin{center}
\includegraphics[width=\textwidth]{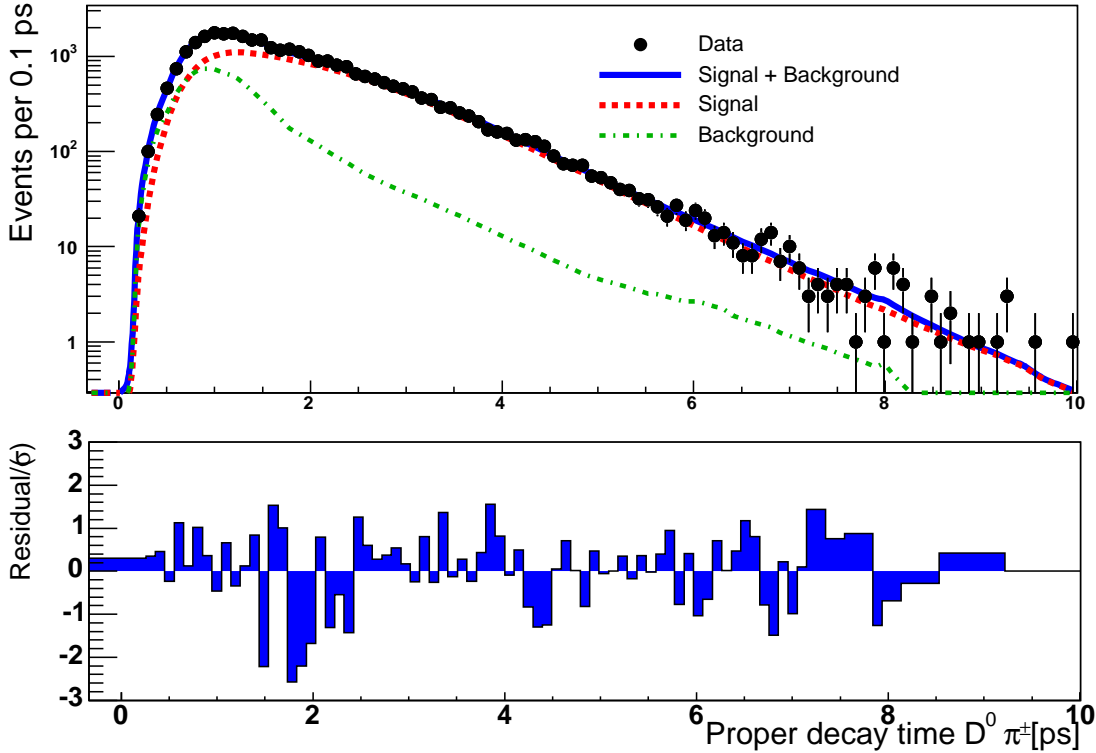}
\end{center}
\caption[Proper decay time fit projection]{This figure
 shows the projection of the lifetime fit onto the data. The signal
 and background components are shown separately (dotted lines) and in
 addition (solid line). The points are data. The lower plot shows the
 residual divided by the error for each bin.\label{fig:lifefit}}
\end{figure*}
\begin{figure}[htpb]
\begin{center}
\includegraphics[scale=0.35]{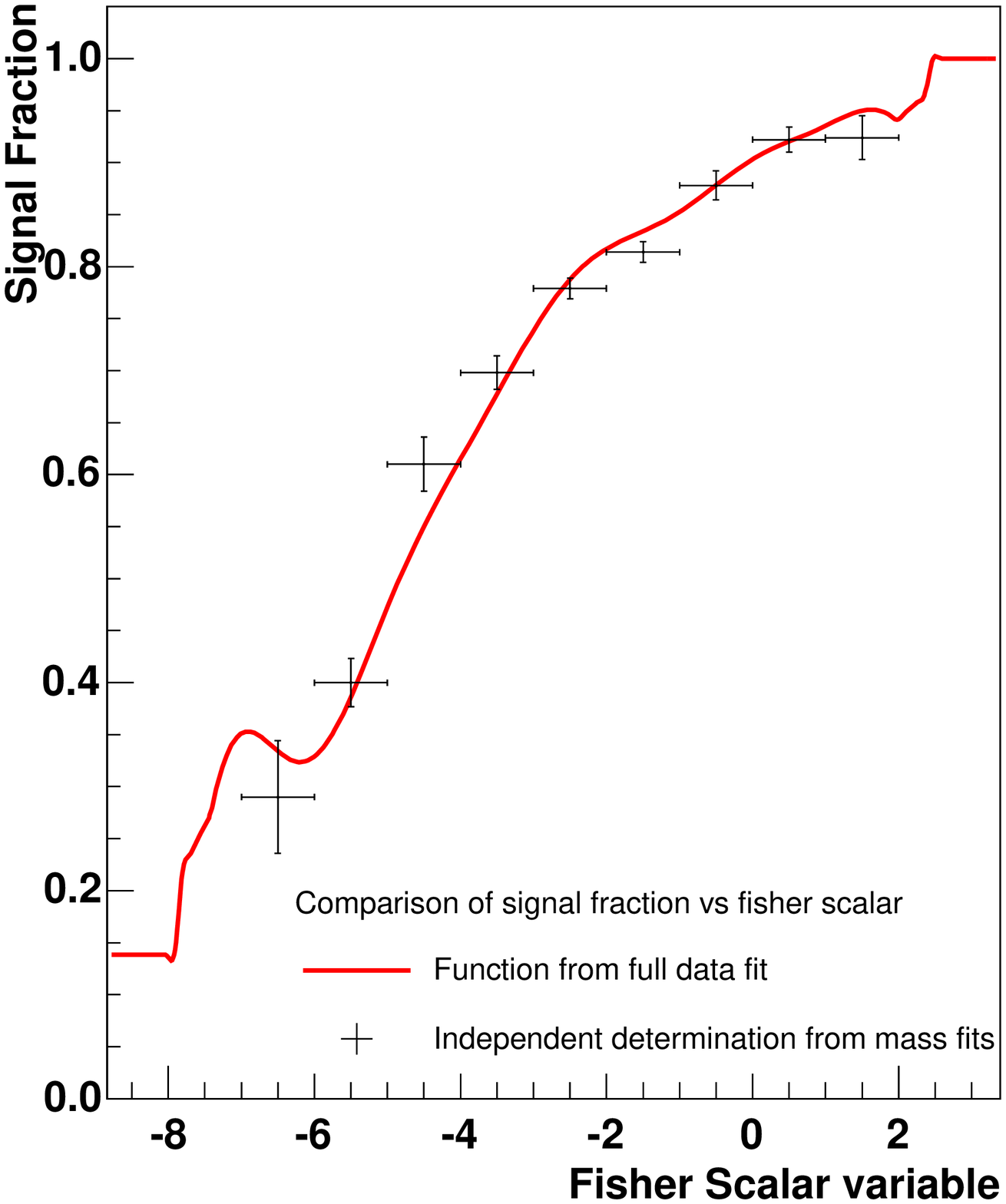}
\end{center}
\caption{Projection of the signal fraction, $f(x)$, as a function of the Fisher
 scalar, determined from the fit (line). The data points are the
 signal fraction determined from mass fits using events only lying
 in that particular bin of $x$.\label{fig:fishdata}}
\end{figure}
\begin{figure}[htbp]
\begin{center}
\includegraphics[scale=0.35]{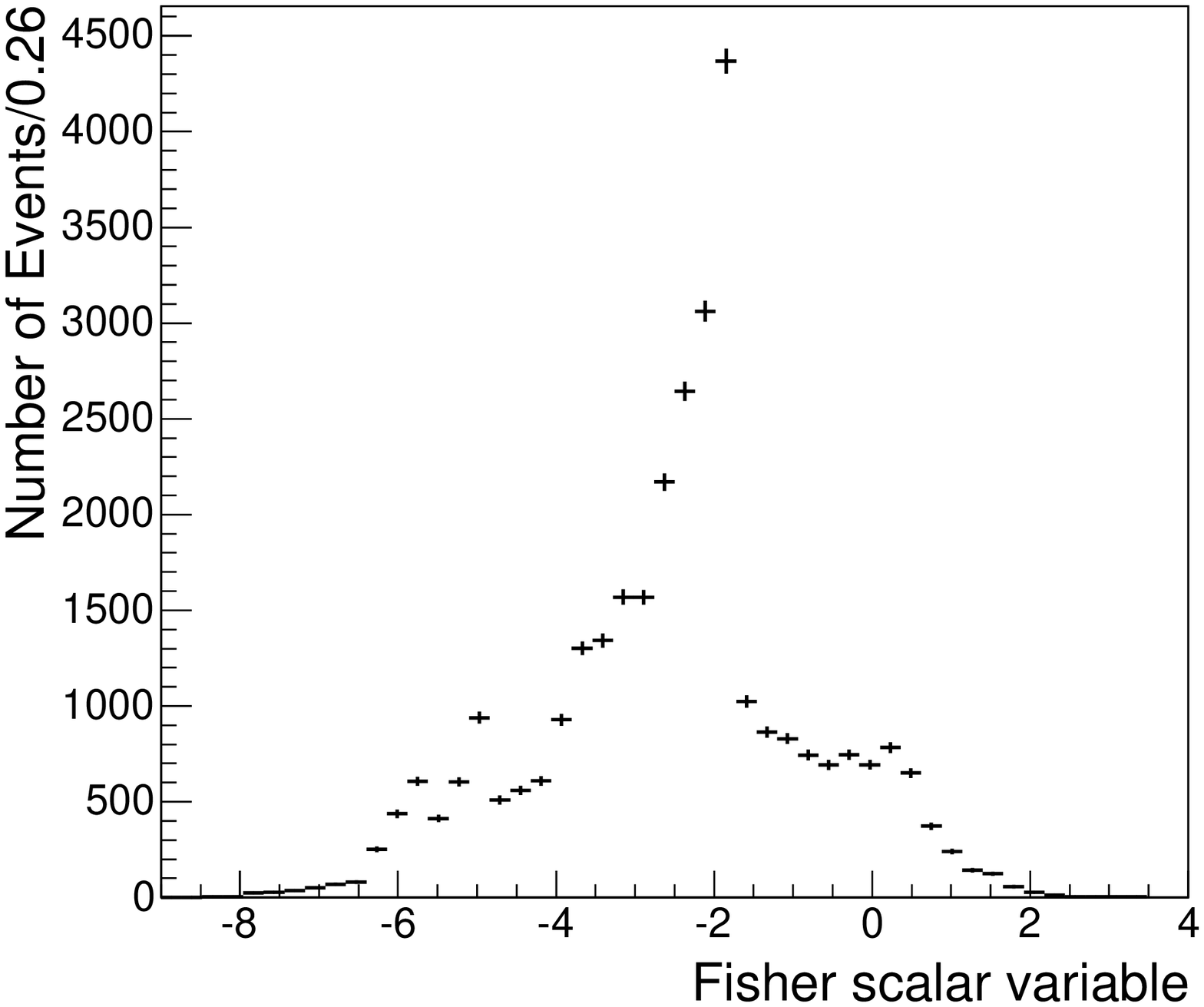}
\end{center}
\caption{The distribution of the Fisher Scalar variable $x$, in data.}
\label{fig:fishdist}
\end{figure}

 The proper decay time fit projection for all events in the fit is shown in
 \figref{fig:lifefit}.  
 The function ,$f(x)$, determined by the fit is shown in
 \figref{fig:fishdata}, and the distribution of the variable $x$
 itself is shown in \figref{fig:fishdist}. To assess how well $f(x)$
 determines the signal fraction, the data with $-7<x<2$ are
 divided into nine bins. A mass fit is performed separately for the
 events in each bin to obtain an independent measure of the signal
 fraction in that bin. For $x$ outside the range $-7<x<2$, there are
 insufficient data to perform a mass fit.
 The signal fractions, as determined by the series of mass fits, are
 overlaid on the function, $f(x)$, in \figref{fig:fishdata}, and there is
 good agreement between the two determinations of signal fraction.

 The fit result for the $B^{-}$ lifetime is
\(
 \un{\tau(B^{-}) = 1.663 \pm 0.023 \mathrm{(stat)}}{ps},
\)
 where the uncertainty has already been scaled by the factor 1.1 as
 discussed in Sec.~\ref{p:fishsummary}. The fit results for all other
 parameters can be found in \Appref{sec:fittab}. The lifetime is only
 weakly correlated to the other fit parameters; the correlation
 coefficient between the lifetime and any other fit parameter is
 always less than $10\%$. The statistical uncertainty on $\tau(B^-)$
 is about twice as large as one would naively expect from dividing the
 fit result by the square-root of the number of signal events,
 $\sigma_{\tau\,\mathrm{naive}} \approx \tau/\sqrt{N_{\mathrm{sg}}} =
 \un{0.011}{ps}$, which usually gives a reasonable estimate for data
 with good proper decay time resolution and small background
 contamination as we have here. As shown in Ref.~\cite{Rademacker:2005ay},
 the cause for the increased uncertainty is the trigger bias,
 specifically the \emph{upper} impact parameter cut in the trigger,
 which leads to a significantly reduced statistical precision per
 event. The size of the effect is consistent with that calculated
 in Ref.~\cite{Rademacker:2005ay}.

%---------------------------------------------------------

\section{Systematic Uncertainties}
\label{sec:Systematics}
In this section we evaluate the systematic uncertainty on our
 measurement from a variety of possible causes. The two dominant
 uncertainties are due to the dependence of the single track finding
 efficiency of the SVT on impact parameter (\secref{s:svtd0}) and the
 correlation between the measured $\tm$ and $\mass$ that we observe in
 background data from the upper sideband.

 We evaluate each uncertainty as follows: for each source of
 uncertainty 1000 samples of simulated data are generated using the
 fast simulation. Each sample contains approximately the same number
 of signal and background candidates as are found in data. The samples are
 generated using a non-standard configuration that simulates the
 effect under consideration; we then extract the $B^-$ lifetime from
 each sample in the same way as we do for data, using the standard PDF
 described in Sec.~\ref{sec:SgBgPDF}. For each source of systematic
 uncertainty, the mean residual, (fitted lifetime $-$ input lifetime),
 averaged over the 1000 samples, is taken as the systematic
 uncertainty. The statistical uncertainty on the mean residual from
 1000 generated samples of simulated data is approximately 0.0007~ps and systematic
 uncertainties of this size or smaller are deemed negligible.

\subsection{The dependence of the single track finding efficiency on
 impact parameter}
\label{s:svtd0}
 The leading source of systematic uncertainty is the parametrization
 of the L2 single track finding efficiency as a function of track
 impact parameter. As described in \secref{sec:removeBiasForSignal},
 we assume that $\effSVT\left(|\dzOff|\right)$ is constant for
 $|\dzOff|<\un{1000}{\mu m}$.  \Figref{f:svtnonflat} shows the efficiency as a function of $|\dzOff|$ in data, and indicates
 that \effSVT\ starts dropping slightly before \un{|\dzOff| = 1000}{\mu
 m}. To obtain a model for the track finding efficiency to use in
 the simulation, we fit the SVT single track finding efficiency as a
 function of $|\dzOff|$ found in data, using the function $\eff(|\dzOff|) =
 p_0 \times G\left(\frac{|\dzOff| - p_1}{p_2}\right)$, where $p_0$, $p_1$,
 and $p_2$ are free parameters and $G(x)$ is the complementary error
 function defined as $G(x)=\frac{2}{\sqrt{\pi}} \int_x^\infty
 \exp(-t^2) dt$.  This fit results in one particular determination of
 the single track finding efficiency shape. We create other SVT
 single-track efficiency distributions, consistent with the data, by
 varying $p_0$, $p_1$ and $p_2$ by the statistical uncertainty of their fitted
 values. Of these distributions we choose the three which we expect to produce the largest biases in the fitted lifetime, i.e., the distributions that have the largest difference in efficiency between $|\dzOff|=0$ and
 $|\dzOff| = \un{1000}{\mu m}$. These three SVT single track
 efficiency functions, one of which is the original fit result itself,
 are represented by the three lines in
 \figref{f:svtnonflat}. The different single track efficiency functions
 are implemented in the simulation by assigning SVT matches with the
 probability determined by the given function. For each of the three
 functions considered, a set of 1000 simulated samples is generated, and fit,
 with the standard PDF that assumes a flat SVT single track efficiency
 for \un{|\dzOff| < 1000}{\mu m}.  The mean lifetime residual from the
 fits to these samples varies from -0.0060 to -0.010~ps, depending
 on the values of $p_0$, $p_1$ and $p_2$ used. To be conservative we
 assign a 0.010~ps systematic uncertainty due to assuming that
 $\effSVT(\dzOff)$ is constant for tracks with impact parameter less than
 \un{1000}{\mu m}.

\subsection{Single track finding efficiency dependence on $p_T$ and $\eta$}
\label{s:svtpteta}
\begin{figure}
\begin{center}
\includegraphics[width=\columnwidth]{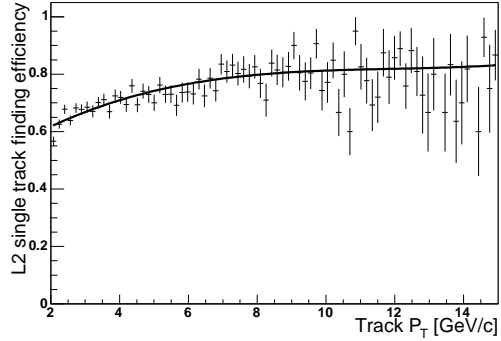}
\end{center}
\caption[Background proper decay time fit projection]{SVT single track
 finding efficiency as a function of track transverse
 momentum. The (line) is a third order polynomial fit to the data (points).\label{fig:pteff}} 
\end{figure}

 The fit also neglects the dependence of $\effSVT$ on $p_T$ and $\eta$
 for tracks that pass the trigger criteria, i.e., with $p_T>2$ GeV/$c$. \Figref{fig:pteff} shows $\effSVT(p_T)$ in data. The line
 through the data represents a fit using a third order polynomial. The efficiency is obtained in a similar manner to \Figref{f:svtnonflat}, where the third track, in the sub-sample of candidates where the other two tracks are sufficient to pass the trigger, is used to determine the efficiency. The $p_T$
 dependence is incorporated into the fast simulation by assigning SVT
 matches based on the probability given by the polynomial function. We determine a systematic uncertainty of 0.006~ps. Similarly, we
 evaluate the effect of the dependence of the SVT single track finding
 efficiency on the track's pseudorapidity and obtain a systematic
 uncertainty of 0.001~ps. The dependence of $\effSVT$ on track $p_T$ and $\eta$ are not large sources of uncertainty as they are not directly related to the proper decay time unlike the impact parameter.

\subsection{Dependence on the impact parameter resolution shape}
 We assume that the impact parameter resolution between
 the offline and online algorithms remains constant as a function of
 impact parameter. As discussed in Sec.~\ref{sec:sigval},
 it has been shown that the technique of sliding the decay vertex is
 insensitive to the actual shape of the resolution as long as the
 shape remains constant.  In the data we do, however, see subtle
 differences, at the level of a few microns, in the mean and width of the resolution as a function of
 impact parameter (\figref{fig:meanswidths},
 \secref{sec:removeBiasForSignal}). To test this effect we incorporate
 such differences, as found in data, into the fast simulation. The
 bias observed due to this is 0.002~ps.

\subsection{Dependence of background observables on mass}
\label{s:bkgcorr}

\begin{figure}[htbp]

\subfigure[]
{\includegraphics[width=\columnwidth]{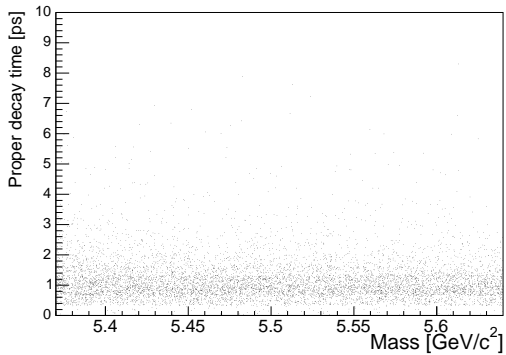}
\label{fig:MLCscat}}
\qquad
\subfigure[]
{\includegraphics[width=\columnwidth]{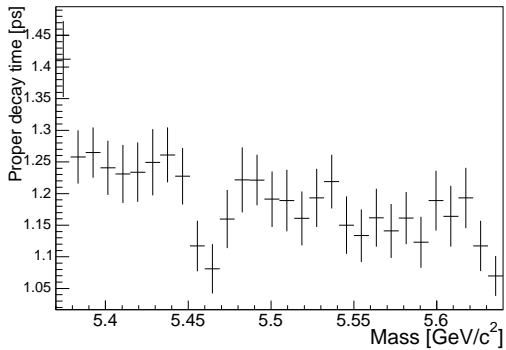}
\label{fig:MLCprof}}
\caption[]{A scatter and a profile plot shows the correlation between the mass and proper decay time of candidates in the upper sideband.\label{fig:tmcorr}}

\end{figure}

 In data we observe a correlation between the measured $\tm$ and
 $\mass$ for background candidates in the upper sideband, which is shown
 in the scatter and profile plot in Fig.~\ref{fig:tmcorr}. We assume
 that this correlation is described well by a linear relationship and
 determine that the mean reconstructed proper decay time of background
 varies by approximately 0.13 ps over a mass range of 0.27
 GeV$/c^2$, which is the mass range used in the lifetime
 fit. The derivation of the PDF assumes that the proper decay time has no dependence on the measured mass of the background
 candidate. To test the effect of neglecting the correlation between $\tm$ and $\mass$ in the PDF
 we extrapolate the same linear correlation for background candidates
 underneath the peak as observed in the sideband. We generate
 simulated data where background candidates are rejected in such a way as
 to introduce a correlation between the mass and proper decay time
 of the candidate, similar to that observed in data. We determine a systematic uncertainty of 0.0083 ps using samples of fast simulation signal and background candidates. This is one of the leading
 sources of systematic uncertainty. It could be reduced in future
 measurements by defining a proper decay time parametrization for
 background that includes dependence on the mass. One possible way to
 do this would be to assume that
 $\mathcal{P}(\tm|\trig,\acc,b,\mass)=\mathcal{P}(t_i^{\dagger}|\trig,\acc,b)$
 where $t_i^{\dagger}=t_i + \beta(\mass-m_0)$,$m_0$ is a central
 mass value, and care is taken to ensure proper normalization.

 In the derivation of the PDF, we also assumed that there was no
 relation between $\acc$ and $\mass$ for background candidates. Candidates in
 the upper sideband are used to calculate $\vect{v}_b$ which is
 necessary to determine the Fisher discriminant. We assume that the
 calculated $\vect{v}_b$ is representative of all background
 candidates. To test the sensitivity of the lifetime result to the
 particular background sample, we repeat the lifetime fit to data but now use candidates with reconstructed mass between $5.5<M_B<5.7$ GeV$/c^2$ to calculate $\vect{v}_b$. There is no change in the
 fitted lifetime for data which demonstrates that there is no
 significant relation between \acc\ and $m_i$ for background candidates.

\subsection{Background proper decay time parametrization}
\label{s:bkgparam}
 To test the reliability of the $y(t)$ parametrization described in
 Sec.~\ref{p:time}, we seek an alternate parametrization of the
 data. We use the sum of two exponentials convoluted with the detector
 resolution. This parametrization of the background is not used in
 the main fit, as the quality of fit to the sideband data is poor. Nonetheless, we can
 generate simulated data where the background proper decay times are generated
 using the sum of two exponential functions with mean lifetimes of
 0.787 ps and 0.0282 ps in the ratio 1:7.3 as found from a fit to
 the sideband. This results in a background proper decay length
 distribution that has similar characteristics to the distribution
 observed in the upper sideband. We fit these simulated data samples
 with the standard PDF. The mean lifetime residual is 0.0027 ps and we
 take this as an estimate of the systematic uncertainty due to the
 background proper decay time parametrization.

\subsection{Silicon Alignment}
\label{s:align}
 To determine the uncertainty due to a possible misalignment of the
 silicon (SVX-II) detector we consider radial shifts in the silicon layers towards and away from the beam pipe, of 50 \mum as has been done in other lifetime
 measurements at the CDF experiment, for example, Ref.~\cite{jpsiLambda}. The shifts in the silicon layers change the measured hit positions of the tracks. To first order, the mis-measurement in track impact parameters are related to a 50 \mum shift in the silicon layers by the relation $\delta (d_0) = 50\cdot \mathrm{sin}(\alpha)$, where $\alpha$ is the angle between the track and the perpendicular to the silicon layer in the transverse plane.
 We recalculate the measured impact parameters in the fast
 simulation containing the misalignment model. The proper decay time of the candidate is recalculated using
 the shifted impact parameter values. Fitting 1000 samples generated in
 this way we find a systematic uncertainty in lifetime measurements
 due to silicon layer misalignment of 0.0013~ps.

\subsection{Detector Resolution Model}
\label{s:resosys}

 In the fit to data, we describe the detector time-resolution with a
 Gaussian of width 0.0087~ps. We estimate the systematic uncertainty due to the chosen resolution model 
 by generating data sets with an alternative resolution function and
 fitting it using the standard PDF. The alternative resolution
 function is described by a sum of three Gaussians with widths 0.0067,
 0.0124, 0.0249~ps and relative fractions 1:0.92:0.04. This resolution
 function derives from a study of prompt $D$ mesons combined with an
 extra track from the primary vertex. We test the effect of this
 alternate resolution using the fast simulation. From the 1000 samples
 of fast simulated data, we find that the mean residual is 0.0010 ps
 and we take this as a systematic uncertainty.

\subsection{Signal composition}
\label{s:cabib}
 We also consider contamination of the signal peak by the decay
 \bdknogap. This decay can appear in the sample if the kaon track is
 reconstructed as a pion and the resulting decay passes the selection
 criteria. Although this is the decay of a charged $B$ meson, the
 proper decay time distribution of this decay mode will be altered as
 the mass has been miscalculated. We use the fast simulation to
 estimate the fraction of \bdk candidates that pass the lower mass
 cut. This information, in conjunction with the relative branching
 fractions of the \btodpi and \bdk decay modes~\cite{PDG}, results in
 the estimate that 3$\%$ of the candidates in the signal peak are actually
 misreconstructed \bdk decays. This fraction is introduced into the
 fast simulation and the effect on the best fit lifetime is
 negligible.

\subsection{Summary}
\label{s:summary}
\begin{table*}

\caption{Summary of systematic uncertainties.\label{tab:systerr}}
\begin{center}
\begin{tabular}{l c}
\hline \hline 
Source of systematic uncertainty & Uncertainty(ps)
\\ \hline
 Track finding efficiency dependence on \dzOff &0.0103\\
Track
finding efficiency dependence on $p_T$ & 0.0060\\ 
Variation in
impact parameter resolution & 0.0020\\ 
Track finding efficiency
dependence on $\eta$ & 0.0010\\ 
Mass-proper decay time
correlation in background & 0.0083\\ 
Background proper decay time
parametrization & 0.0027\\ 
Silicon alignment & 0.0013\\ 
Transformation of \acct\ to scalar variable & 0.0013\\ 
Detector
resolution model & 0.0010\\ 
Signal composition & negligible\\
\hline
Total systematic uncertainty & 0.015\\\hline\hline
\end{tabular}
\end{center}
\end{table*}
 A list of systematic uncertainties is given in
 \tabref{tab:systerr}.  We combine the uncertainties in quadrature to
 find a total systematic uncertainty of \un{0.015}{ps}, which is smaller than the statistical uncertainty of \un{0.023}{ps}. The leading sources of systematic uncertainty are related to the
 details of the SVT single track finding efficiency, and the
 correlation in background between the reconstructed proper decay time
 and mass.
 Neither of these are irreducible and, should the systematic
 uncertainty become a limiting factor in future measurements, it
 should be possible to improve them significantly. A more detailed
 description of the SVT track finding efficiency, which can be
 obtained from data, can be incorporated into the fit to reduce the
 leading systematic error.  Similarly, the correlation between the
 mass and proper decay time in background candidates can be incorporated
 into a future version of this technique to reduce the second largest
 contribution.

\section{Result and Conclusion}
\label{sec:Conclusion}

 We introduce a simulation-independent method for measuring lifetimes
 in event samples where the selection criteria bias the proper decay time
 distribution. We apply it to measure the $B^-$ lifetime in data
 collected by the hadronic $B$ trigger at CDF, which selects events with
 displaced tracks and thus biases the measured proper decay time distribution.

 In previous analyses, the trigger bias has been corrected for using an efficiency function obtained from Monte Carlo simulation.  This simulation dependence can
 be a significant source of systematic uncertainty. A recent example
 is the measurement of the $\Lambda_b$ lifetime in the hadronic decay
 channel $\Lambda_b \to \Lambda_c \pi$ at CDF, which found
 $\tau(\Lambda_b)=1.401\pm 0.046$ (stat) $\pm 0.035$ (syst)~ps~\cite{mumford}. 
The systematic uncertainty in this
 measurement is almost entirely due to the simulation
 dependence. While currently smaller than the statistical uncertainty, the systematic uncertainty could limit the precision in future, higher statistics measurements.

 The method introduced here removes the simulation dependence by
 replacing the global efficiency function with candidate-by-candidate
 efficiency functions that can be calculated analytically from the
 event data, without recourse to simulation.
 We test the method extensively with simulated data, and finally apply
 it to measure the lifetime of the $B^-$ meson, $\tau(B^-)$, using
 $23900 \pm 200$ \btodpi candidates, where \dtopiknogap , collected by CDF's hadronic $B$ trigger in \un{1}{fb^{-1}} of data. We extract
 $\tau(B^-)$ from the data without input from simulation. We measure $\tau(B^-)=1.663 \pm 0.023$ (stat) $\pm 0.015$ (syst) ps. This result is in good agreement with the world average of $1.638 \pm
 0.011$~ps~\cite{PDG}. This technique generalizes easily to other decay channels, as we have
 demonstrated in Sec.~\ref{sec:FullSim}. It can be applied to any
 situation where the trigger or other selection criteria bias the
 proper decay time distribution of the reconstructed data.

\section{Acknowledgments}
\label{sec:sec9}

We thank the Fermilab staff and the technical staffs of the participating institutions for their vital contributions. This work was supported by the U.S. Department of Energy and National Science Foundation; the Italian Istituto Nazionale di Fisica Nucleare; the Ministry of Education, Culture, Sports, Science and Technology of Japan; the Natural Sciences and Engineering Research Council of Canada; the National Science Council of the Republic of China; the Swiss National Science Foundation; the A.P. Sloan Foundation; the Bundesministerium f\"ur Bildung und Forschung, Germany; the World Class University Program, the National Research Foundation of Korea; the Science and Technology Facilities Council and the Royal Society, UK; the Institut National de Physique Nucleaire et Physique des Particules/CNRS; the Russian Foundation for Basic Research; the Ministerio de Ciencia e Innovaci\'{o}n, and Programa Consolider-Ingenio 2010, Spain; the Slovak R\&D Agency; and the Academy of Finland.

%------------------------------------------------------------

\appendix
\section{Factorizing the PDF}
\label{sec:Derivation}
 This appendix details the factorization of the PDF term
 $\mathcal{P}(s,\tm,\mass,\trk,\acc;\tau|\trig)$ which describes the
 probability that we observe an event with given values of
 $\tm$, $\mass$, $\acc$, and $\trk$. Although there are a number of
 ways to factorize the expression in \equationref{eq:fullPDF:sb} we
 aim to find a final form that includes the factor $
 \mathcal{P}(t_i;\tau|\trig,\acc,s) \times
 \mathcal{P}(\trk|\trig,\acc,\tm,s) $ as a parametrization for this
 factor, given in \equationref{eq:withSVTProb2} is well understood.
 We make use of the following relation
\begin{equation}
\label{e:pab}
 \mathcal{P}(A,B)=\mathcal{P}(A)\mathcal{P}(B|A)
 =\mathcal{P}(B)\mathcal{P}(A|B).
\end{equation}
 We only explicitly write the dependence of the PDF on the observables. Using \equationref{e:pab},
 $\mathcal{P}(s,\tm,\mass,\trk,\acc;\tau|\trig) $ can be split into
 two factors:
\begin{equation}
\label{e:app1}
\begin{split}
 \mathcal{P}&(s,\tm,\mass,\trk,\acc;\tau|\trig) = 
\\ &   \mathcal{P}(s,\tm,\acc;\tau|\trig)
\\ & \times \mathcal{P}(\mass,\trk|\trig,\acc,\tm,s;\tau).
\end{split}
\end{equation}
 The last factor, $ \mathcal{P}(\mass,\trk|\trig,\acc,\tm,s;\tau)$ is
 concerned with the probability of observing a particular mass and
 track configuration. This can be factorized further, again using
 \equationref{e:pab}:
\begin{equation}
\label{e:app2}
\begin{split}
\mathcal{P}&(\mass,\trk|\trig,\acc,\tm,s;\tau) = 
\\ &   \mathcal{P}(\mass|\trig,\acc,\tm,s;\tau)
\\ & \times \mathcal{P}(\trk|\mass,\trig,\acc,\tm,s;\tau).
\end{split}
\end{equation}
 The measured mass is independent of the mean lifetime and of the
 efficiency function and we make the assumption that it is independent
 of the measured proper decay time. With these simplifications we can say that
\begin{equation}
\label{e:app2a}
\mathcal{P}(\mass|\trig,\acc,\tm,s;\tau)=\mathcal{P}(\mass|\trig,s).
\end{equation}
 The track configuration is independent of the mean lifetime and the
 mass and therefore we can make the simplification
 $\mathcal{P}(\trk|\mass,\trig,\acc,\tm,s;\tau)=\mathcal{P}(\trk|\trig,\acc,\tm,s)$. Substituting
 this and \equationref{e:app2a} into \equationref{e:app2} leads to
\begin{equation}
\label{e:app2b}
\begin{split}
  \mathcal{P}&(\mass,\trk|\trig,\acc,\tm,s;\tau) = 
\\ & \mathcal{P}(\mass|\trig,s)
\;\mathcal{P}(\trk|\trig,\acc,\tm,s).
\end{split}
\end{equation}
 The remaining factor in \equationref{e:app1} is factorized further using
 \equationref{e:pab}:
\begin{equation}
\label{e:app3}
\begin{split}
\mathcal{P}&(s,\tm,\acc;\tau|\trig) =
\\ &    \mathcal{P}(\tm;\tau|\trig,\acc,s)
  \;\mathcal{P}(s,\acc|\trig). 
\end{split}
\end{equation}
 The first factor on the right hand side corresponds to the first factor in \equationref{eq:withSVTProb2}. Applying
 \equationref{e:pab} one more time we find
 $\mathcal{P}(s,\acc|\trig)=\mathcal{P}(s|\acc,\trig)\mathcal{P}(\acc|\trig)$.
 Therefore,
\begin{equation}
\label{e:app3a}
\begin{split}
\mathcal{P}(s,\tm,&\acc;\tau|\trig) = 
 \\ & \mathcal{P}(\tm;\tau|\trig,\acc,s)
 \\ & \times \mathcal{P}(\acc|\trig)
  \;\mathcal{P}(s|\acc,\trig).
\end{split}
\end{equation}
 Substitution of \equationref{e:app3a} and \equationref{e:app2b} into
 \equationref{e:app1} leads to
\begin{equation}
\label{eq:fullpdfInDerivationAppendix}
\begin{split}
  \mathcal{P} & \left(s, \tm, \mass, \trk, \acc;\tau | \trig \right)=
\\ & \mathcal{P}(\tm;\tau|\trig, \acc, s)
\\ & \times \mathcal{P}(\trk |\trig, \acc, \tm, s) 
         \; \mathcal{P}(\acc |\trig)
\\ & \times \mathcal{P}(\mass|\trig, s)
         \; \mathcal{P}(s|\acc,\trig),
\end{split}
\end{equation}
 which is the same as the expression given in \equationref{eq:fullpdf}
 in Sec.~\ref{sec:SgBgPDF}.

 There are other ways to factorize the PDF. One of
 particular interest is a parameterization that depends on the overall
 signal fraction rather than the event-by-event signal probability
 used here. This can be obtained by replacing
 $\mathcal{P}(s|\acc,\trig)$ with $\mathcal{P}(s)\mathcal{P}(\acc|
 s,\trig)$ (and equivalently for the background terms). $P(s)$, often
 written as $f_s$, is the overall signal fraction and $P(b) = 1-P(s)$
 is the background fraction. This PDF differs from the one we use by an
 overall factor $P(\acc | \trig)$ which does not affect the maximum of
 the likelihood function. Our choice of PDF is driven by the ease of parameterization of the required function. It is easier to parameterize the smoothly varying candidate-by-candidate signal probability (see \figref{fig:fishdata}, $P(s|\acc,\trig)$), rather than parameterise the fine structure observed in the fisher scalar distribution (see \figref{fig:fishdist}, $P(\acc|s,\trig)$).

\section{Characterizing the efficiency function by a vector of variables}
\label{sec:FisherAppendix}
In order to parametrize the term $\mathcal{P}(s|\acc, \trig)$, which
 arises in the PDF in Eq. (\ref{eq:fullpdf}) we have chosen to use
 Fisher Discriminant Analysis to characterize each \acc\ by a scalar variable
 $x_i$ (as described in Sec.~\ref{s:scalar}). To use the Fisher
 Discriminant method we need to construct a vector $\vect{v}_i$,
 whose components describe the efficiency function, $\acc$. How this
 vector is obtained is described here.

  The vector $\vect{v}_i$ contains a series of variables: $v_1$, $v_2$..., $v_n$ . Each variable should describe a property of the efficiency
  function in a way that allows comparison of one candidate's efficiency
  function with that of another. In Sec.~\ref{s:calceff} we showed that the
  efficiency function can be written as 
\begin{equation}
\label{eq:accdesc}
\begin{split}
\lefteqn{\acc =} & \\ &
 \intervSum\left[
       \theta(t-t_{\mathrm{min}\;k_i}) 
    - \theta(t-t_{\mathrm{max}\;k_i})
\right] H_{k_i}(\effSVT),
\end{split}
\end{equation}
 where $H(\effSVT)$ is either $\effSVT^2$, $2\effSVT^2 - \effSVT^3$ or
 $3\effSVT^2-2\effSVT^3$, depending on whether there were one, two or three
 track pairs that could have passed the trigger.

From inspection of Eq.(\ref{eq:accdesc}), a single efficiency function can be uniquely defined by a series of variables that are $t_{\mathrm{min}\;k_i}$, $t_{\mathrm{max}\;k_i}$ and the value of $H_k(\effSVT)$. However, this is not a useful description for comparing one efficiency function to the next because the number of intervals, and hence the number of variables required to describe the efficiency function, varies from one candidate to the next. 

For all candidates the efficiency function is defined for proper decay times in the range 0--10 ps. Another way to construct $\vect{v}_i$ would be to bin the efficiency into $n$ equal bins of time and take the mean value of $H_k(\effSVT)$ in each bin as the elements $v_1$, $v_2 ...$, $v_n$. In this way, the value of any particular element of $\vect{v}_i$ for one candidate can be compared to the same element of $\vect{v}_i$ for another candidate. This method is also problematic as the $\effSVT$ is a floating parameter and the Fisher Discriminant (and, hence, the observable $x_i$) cannot be recalculated at each iteration of the likelihood minimization.

We take an approach that allows construction of $\vect{v_i}$ so that its elements can be compared across candidates without requiring knowledge of the value of $\effSVT$. The efficiency function can be re-written in the form 
\begin{equation}
\label{eq:accdesc2}
\begin{split}
\acc
 = & \mathcal{A}_a\cdot \effSVT^2 + \mathcal{A}_b\cdot (2\effSVT^2-\effSVT^3) 
\\ & + \mathcal{A}_c\cdot (3\effSVT^2-2\effSVT^3),
\end{split}
\end{equation}
where,
\begin{equation}
\mathcal{A}_a =\!\!\!\!\!\! \sum\limits_{\parbox{6.5em}{\center\scriptsize
 \mbox{}\vspace{-4ex}\\$j$=all intervals with\vspace{-0.5ex}\newline $H(\effSVT)=\effSVT^2$}}
\!\!\!\!\!\!\!\!
\left[
\theta(t-t_{\mathrm{min}\;j}) - \theta(t-t_{\mathrm{max}\;j})\right], 
\end{equation}
and corresponding terms for $\mathcal{A}_b$ and $\mathcal{A}_c$. The value of the functions $\mathcal{A}$ at any time are either 0 or 1. Writing the efficiency function this way splits it into three sections, dependent on the value of $H(\effSVT)$. Comparing  $\mathcal{A}_a$ from one candidate to the next allows comparison of the efficiency function arising from the parts where there was only one track pair available to pass the trigger. To construct $\vect{v_i}$ we bin each of the  $\mathcal{A}$ functions into 20 bins as a function of proper decay time. The value of $v_1 ... v_{20}$ are the values of $\mathcal{A}_a$ in each bin.  Nominally the value in any given bin is either 0 or 1, however, where the efficiency turns on or off within the bin an intermediate value is taken to represent the mean efficiency in that bin. Similarly the values of $v_{21} ...v_{40}$ are the values of $\mathcal{A}_b$ in each bin and $v_{41}...v_{60}$ are the values of $\mathcal{A}_c$. By splitting the efficiency function into three parts, dependent on the form of $H_k(\effSVT)$, we have found a vectorial representation of the efficiency function that is independent of the absolute value of $\effSVT$ and that allows comparison of \acc\ between different candidates. We now have a prescription for converting \acc\ into $\vect{v_i}$ for each candidate. The mean $\vect{v}$ for background events, $\vect{v_b}$, can be found from averaging the $\vect{v}_i$ for candidates with 5.37 GeV/$c^2$ $>m_B$, i.e.,
\begin{equation}
\vect{v_b} = \frac{\sum_{m_B>5.37 GeV/c^2} \vect{v}_i}{\sum_{m_B>5.37 GeV/c^2} 1}.
\end{equation}
To determine $\vect{v_s}$ we first determine $\vect{v_r}$, which is the average of the $\vect{v}_i$ for events that have mass in the range 5.25$<m_B<$5.32 GeV/$c^2$. As this region contains both signal and background events, $\vect{v_s}$ can be determined from $\vect{v_r}$ by subtracting the appropriate fraction of $\vect{v_b}$. This fraction is determined from a fit to the mass distribution. Having determined $\vect{v_b}$ and $\vect{v_s}$, the direction $\vect{w}$, and therefore $x_i$ can be determined using Fisher Discriminant Analysis~\cite{Fisher1}.

\section{A simpler PDF}
\label{sec:simplerPdf}
A lot of the complexity of the method presented here results directly
or indirectly from the tight upper impact parameter cut applied by the
two track trigger. In situations where this upper impact parameter cut
is significantly looser, or ideally where no such cut is applied at
all, one would not only benefit from a higher statistical precision
for each candidate~\cite{Rademacker:2005ay}, but would also be able to
employ a significantly simpler version of the method as
outlined below. In this simpler version
\begin{itemize}
 \item the dependence on $\effSVT$ can be removed, 
 \item under many circumstances, there is no need to use the Fisher
 discriminant.
\end{itemize}
While we did not choose this approach for reasons specific to the CDF II detector trigger (as discussed below), it is summarized here for the benefit of
potential users of this method at other experiments.
\subsection*{Removing the dependence on \effSVT}
As described in \secref{varSVTeff}, if the track-finding efficiency is decay time independent, one can base a
fit on the PDF \emph{given} that a certain track combination has been
reconstructed and seen by the trigger. Given that a certain track combination has been found, the
trigger efficiency at a certain decay time is either $1$ (passes cuts)
or $0$ (fails), independent of \effSVT. With this,
the signal PDF given in \equationref{eq:withSVTProb} reduces to:
\begin{widetext}
\begin{equation}
\label{eq:simplerPDFnoEff}
\begin{split}
 \mathcal{P}(t_i;\tau|\trig,\acct,s)  =
    \frac{  
           \frac{1}{\tau}e^{\frac{-t_i}{\tau} + \frac{1}{2} \frac{\sigma_t^2}{\tau^2}}
           \Freq\left( \frac{t_i}{\sigma_t} - \frac{\sigma_t}{\tau}\right)
   }{
   \intervSum\!\!\!
    \left[
           - e^{\frac{-t}{\tau} + \frac{1}{2} \frac{\sigma_t^2}{\tau^2}}
           \Freq\left( \frac{t}{\sigma_t} - \frac{\sigma_t}{\tau}\right)
           + \Freq\left(\frac{t}{\sigma_t}\right)
    \right]_{t=t_{\mathrm{min\; {k_i}}}}^{t=t_{\mathrm{max\; {k_i}}}}
   }.
\end{split}
\end{equation}
\end{widetext}
 This approach, which is independent of \effSVT, is valid whenever the
 track-finding efficiency is independent of the decay time for all
 tracks in the candidate. Despite the drop of the SVT track finding
 efficiency beyond $|\dzOff| > \un{1}{mm}$, this approach could, in
 principle, be used in the data analyzed in this paper if we applied a
 fiducial cut of $|\dzOff| < \un{1}{mm}$ (where the SVT efficiency is
 effectively constant) to \emph{all} tracks in the decay (this cut
 would of course need to be reflected in the efficiency function
 calculation). This is a significantly harsher requirement than that
 of the trigger, which requires only two out of three tracks to have 0.12$
 < |\dzSVT| < \un{1}{mm}$, allowing one of the tracks to have $|\dzSVT|>
 \un{1}{mm}$. We studied this option and found that the loss in
 statistical precision due to the additional cut is too large, mainly
 because of the effects discussed in
 Ref.~\cite{Rademacker:2005ay}. This simpler approach would however be
 suitable in a situation where the track-finding efficiency is
 constant over a larger range than for the SVT.

\subsection*{Removing the need for a Fisher Discriminant}
 If the dependence on \effSVT\ has been removed as described above,
 and in addition there is no variable upper proper decay time cut (no
 upper impact parameter cut), the candidate-by-candidate \acct\ is
 fully determined by one single parameter, the decay time
 $t_{\mathrm{min}}$ where the acceptance ``turns on'', i.e., above
 which the decay is accepted. Remembering that the motivation for
 introducing the Fisher discriminant was to translate the efficiency
 function into a single number, this would clearly be unnecessary, as
 \acct\ is already fully described by a single number,
 $t_{\mathrm{min}}$. The factor $P(s|\acc)$ can then be replaced by
 $P(s | t_{\mathrm{min}})$, with $P(b | \acc) = P(b |
 t_{\mathrm{min}}) = 1 - P(s|t_{\mathrm{min}}).$ There is now no need
 for the Fisher scalar variable although the PDF term still requires a
 description of the signal fraction as a function of $t_{min}$.

\subsection*{Even simpler: Redefining t=0}
 Finally, in the case where there is no upper lifetime cut
 (i.e. $t_{\mathrm{max}} = \infty$), and the lower lifetime cut is
 hard enough to satisfy $t_{\mathrm{min}} \gg \sigma_t$, all the
 ``F-terms'' in \equationref{eq:simplerPDFnoEff} are 1, and the
 equation reduces to
\begin{equation}
P(t)= 
      \frac{1}{\tau} e^{-(t-t_{\mathrm{min}})/\tau} 
\end{equation}
which is equivalent to an event-by-event re-definition of $t=0$, as
used by DELPHI in Ref.~\cite{Adam:1995mb}.

\section{Full fit results}

\label{sec:fittab}
\begin{table}[htbp]
%\sans
\caption[Best fit mass parameters and uncertainties]{Summary of best fit mass parameters and uncertainties.}  \label{tab:massfitres}
\begin{center}
\begin{tabular}{ccc}
\hline \hline Parameter &Best fit & Uncertainty\\ \hline 
$m_1$  [GeV/$c^2$]  & 5.2762   & $\pm$0.0004 \\ 
$m_2$  [GeV/$c^2$]& 5.2711  & $\pm$0.0025   \\ 
$\sigma_1$ [GeV/$c^2$] & 0.0247  & $\pm$0.0033 \\ 
$\sigma_2$ [GeV/$c^2$ ]& 0.0138  & $\pm$0.0010  \\ 
$f_1$ & 0.481 & $\pm$0.13  \\ 
$f_s$ & 0.741       & $\pm$0.0050 \\
$\alpha$ [(GeV/$c^2$)$^{-1}$]& -0.1658        & $\pm$0.0035  \\ \hline \hline

\end{tabular}
\end{center}

\end{table}

\begin{table}[htbp]
%\sans
\caption[Best fit single track finding efficiency and uncertainties. ]{Summary of best fit efficiency parameters and uncertainties. The three periods correspond to the changes in the SVT described in Sec.~\ref{sec:trigselec}.} \label{tab:effres}
\begin{center}
\begin{tabular}{lcc}
\hline \hline Efficiency parameter &Best fit & Uncertainty\\ \hline
Signal Period 1& 0.488   & $\pm$0.033  \\ 
Signal Period 2& 0.656   & $\pm$0.009  \\ 
Signal Period 3& 0.725  & $\pm$0.006      \\ 
Background Period 1  & 0.496 &$\pm$0.064   \\
Background Period 2 & 0.502 & $\pm$0.019  \\ 
Background Period 3 & 0.560       & $\pm$0.017 \\ \hline \hline

\end{tabular}
\end{center}

\end{table}

\begin{table}[htbp]
%\sans
\caption[Best fit Background proper decay time parameters and uncertainties]{Summary of best fit background proper decay time parameters and uncertainties. The $ct$ value represents the points where the background proper decay time distribution is sampled as defined by $t_j$ in Eq. (\ref{e:bkglife}).} \label{tab:bkgliferes}
\begin{center}
\begin{tabular}{cccc}
\hline \hline 
\parbox{3.5em}{Backg. Parameter}
 &  $ct_j$ ($\mu$m) & Best fit & Uncertainty\\ \hline 
$a_1$ &   0         & 10.80& $\pm$0.39  \\ 
$a_2$ &   146.9        &  7.08& $\pm$0.06  \\
$a_3$ &   322.6        &  4.79 & $\pm$0.04      \\
$a_4$  &   532.7        &  2.73 & $\pm$0.04      \\
 $a_5$&   783.9        &  1.29 & $\pm$0.07      \\ 
$a_6$  &   1084.3        &  1.28 & $\pm$0.10      \\
$a_7$ &   1443.5        &  -1.19 & $\pm$0.19      \\
 $a_8$ &   1873.1        &   -1.93& $\pm$0.29      \\
 $a_9$ &   2386.7        &   -2.73& $\pm$0.47      \\
 $a_{10}$ &  3000         &   -7.16& $\pm$2.87      \\ \hline \hline 

\end{tabular}

\end{center}
\end{table}

\begin{table}[htbp]
%\sans
\caption[Best fit signal fraction as a function of Fisher scalar and uncertainties]{Summary of best fit signal fraction, and uncertainties, as a function of Fisher scalar. The values of Fisher scalar gives the mid point of each bin used by the Lagrange interpolating polynomial function as described in Sec.~\ref{p:fish}.} \label{tab:fishres}
\begin{center}
\begin{tabular}{cccc}
\hline \hline \parbox{3.5em}{Fisher parameter} & \parbox{4em}{Fisher scalar, $x$}  & \parbox{4em}{Best fit signal fraction} & Uncertainty\\ \hline
$x_1$ &   -8.35         & 0.139& $\pm$0.072  \\ 
$x_2$ &   -7.05        &  0.273& $\pm$0.071  \\ 
$x_3$ &   -6.19        &  0.333 & $\pm$0.030      \\
$x_4$ &   -5.32        &  0.379 & $\pm$0.014     \\ 
$x_5$ &   -4.45        &  0.535 & $\pm$0.011      \\
$x_6$ &   -3.59        &  0.657 & $\pm$0.008      \\
$x_7$ &   -2.73        &  0.768 & $\pm$0.006      \\
$x_8$ &   -1.86        &  0.825& $\pm$0.005      \\ 
$x_9$ &   -1.00       &   0.860& $\pm$0.007      \\ 
$x_{10}$ &  -0.13        &  0.907& $\pm$0.007      \\ 
$x_{11}$ &   0.73       &   0.937& $\pm$0.011     \\ 
$x_{12}$&   1.60       &   0.941& $\pm$0.034      \\ 
$x_{13}$&    2.89      &   1.00& $\pm$0.045      \\ \hline \hline
\end{tabular}
\end{center}

\end{table}

%\bibliography{references}

\begin{thebibliography}{31}
\expandafter\ifx\csname natexlab\endcsname\relax\def\natexlab#1{#1}\fi
\expandafter\ifx\csname bibnamefont\endcsname\relax
  \def\bibnamefont#1{#1}\fi
\expandafter\ifx\csname bibfnamefont\endcsname\relax
  \def\bibfnamefont#1{#1}\fi
\expandafter\ifx\csname citenamefont\endcsname\relax
  \def\citenamefont#1{#1}\fi
\expandafter\ifx\csname url\endcsname\relax
  \def\url#1{\texttt{#1}}\fi
\expandafter\ifx\csname urlprefix\endcsname\relax\def\urlprefix{URL }\fi
\providecommand{\bibinfo}[2]{#2}
\providecommand{\eprint}[2][]{\url{#2}}

\bibitem[{\citenamefont{Cabibbo}(1963)}]{CKM1}
\bibinfo{author}{\bibfnamefont{N.}~\bibnamefont{Cabibbo}},
  \bibinfo{journal}{Phys. Rev. Lett.} \textbf{\bibinfo{volume}{10}},
  \bibinfo{pages}{531} (\bibinfo{year}{1963}).

\bibitem[{\citenamefont{Kobayashi and Maskawa}(1973)}]{CKM2}
\bibinfo{author}{\bibfnamefont{M.}~\bibnamefont{Kobayashi}} \bibnamefont{and}
  \bibinfo{author}{\bibfnamefont{T.}~\bibnamefont{Maskawa}},
  \bibinfo{journal}{Prog. Theor. Phys.} \textbf{\bibinfo{volume}{49}},
  \bibinfo{pages}{652} (\bibinfo{year}{1973}).

\bibitem[{\citenamefont{Bigi {\em et~al.}}(1997)\citenamefont{Bigi, Shifman,
  and Uraltsev}}]{bigishifman}
\bibinfo{author}{\bibfnamefont{I.}~\bibnamefont{Bigi}},
  \bibinfo{author}{\bibfnamefont{M.}~\bibnamefont{Shifman}}, \bibnamefont{and}
  \bibinfo{author}{\bibfnamefont{N.}~\bibnamefont{Uraltsev}},
  \bibinfo{journal}{Annu. Rev. Nucl. Part. Sci.} \textbf{\bibinfo{volume}{47}},
  \bibinfo{pages}{591} (\bibinfo{year}{1997}).

\bibitem[{\citenamefont{Tarantino}(2004)}]{tarantino}
\bibinfo{author}{\bibfnamefont{C.}~\bibnamefont{Tarantino}},
  \bibinfo{journal}{Eur. Phys. J} \textbf{\bibinfo{volume}{C33}},
  \bibinfo{pages}{S895} (\bibinfo{year}{2004}).

\bibitem[{\citenamefont{Gabbiani {\em et~al.}}(2003)\citenamefont{Gabbiani,
  Onischenko, and Petrov}}]{Gabbiani1}
\bibinfo{author}{\bibfnamefont{F.}~\bibnamefont{Gabbiani}},
  \bibinfo{author}{\bibfnamefont{A.~I.} \bibnamefont{Onischenko}},
  \bibnamefont{and} \bibinfo{author}{\bibfnamefont{A.~A.}
  \bibnamefont{Petrov}}, \bibinfo{journal}{Phys. Rev. D}
  \textbf{\bibinfo{volume}{68}}, \bibinfo{pages}{114006}
  (\bibinfo{year}{2003}).

\bibitem[{\citenamefont{Gabbiani {\em et~al.}}(2004)\citenamefont{Gabbiani,
  Onischenko, and Petrov}}]{Gabbiani2}
\bibinfo{author}{\bibfnamefont{F.}~\bibnamefont{Gabbiani}},
  \bibinfo{author}{\bibfnamefont{A.~I.} \bibnamefont{Onischenko}},
  \bibnamefont{and} \bibinfo{author}{\bibfnamefont{A.~A.}
  \bibnamefont{Petrov}}, \bibinfo{journal}{Phys. Rev. D}
  \textbf{\bibinfo{volume}{70}}, \bibinfo{pages}{094031}
  (\bibinfo{year}{2004}).

\bibitem[{\citenamefont{Lenz}(2008)}]{Lenz:2008zz}
\bibinfo{author}{\bibfnamefont{A.~J.} \bibnamefont{Lenz}},
  \bibinfo{journal}{AIP Conf. Proc.} \textbf{\bibinfo{volume}{1026}},
  \bibinfo{pages}{36} (\bibinfo{year}{2008}), \eprint{0802.0977}.

\bibitem[{\citenamefont{Beneke {\em et~al.}}(2002)\citenamefont{Beneke,
  Buchalla, Greub, Lenz, and Nierste}}]{Beneke:2002rj}
\bibinfo{author}{\bibfnamefont{M.}~\bibnamefont{Beneke}},
  \bibinfo{author}{\bibfnamefont{G.}~\bibnamefont{Buchalla}},
  \bibinfo{author}{\bibfnamefont{C.}~\bibnamefont{Greub}},
  \bibinfo{author}{\bibfnamefont{A.}~\bibnamefont{Lenz}}, \bibnamefont{and}
  \bibinfo{author}{\bibfnamefont{U.}~\bibnamefont{Nierste}},
  \bibinfo{journal}{Nucl. Phys.} \textbf{\bibinfo{volume}{B639}},
  \bibinfo{pages}{389} (\bibinfo{year}{2002}), \eprint{hep-ph/0202106}.

\bibitem[{\citenamefont{Franco {\em et~al.}}(2002)\citenamefont{Franco, Lubicz,
  Mescia, and Tarantino}}]{Franco:2002fc}
\bibinfo{author}{\bibfnamefont{E.}~\bibnamefont{Franco}},
  \bibinfo{author}{\bibfnamefont{V.}~\bibnamefont{Lubicz}},
  \bibinfo{author}{\bibfnamefont{F.}~\bibnamefont{Mescia}}, \bibnamefont{and}
  \bibinfo{author}{\bibfnamefont{C.}~\bibnamefont{Tarantino}},
  \bibinfo{journal}{Nucl. Phys.} \textbf{\bibinfo{volume}{B633}},
  \bibinfo{pages}{212} (\bibinfo{year}{2002}), \eprint{hep-ph/0203089}.

\bibitem[{\citenamefont{Abulencia {\em et~al.}}(2006{\natexlab{a}})}]{bsmixing}
\bibinfo{author}{\bibfnamefont{A.}~\bibnamefont{Abulencia}} \bibnamefont{{\em
  et~al.}} (\bibinfo{collaboration}{CDF Collaboration}),
  \bibinfo{journal}{Phys. Rev. Lett.} \textbf{\bibinfo{volume}{97}},
  \bibinfo{pages}{242003} (\bibinfo{year}{2006}{\natexlab{a}}).

\bibitem[{\citenamefont{Acosta {\em et~al.}}(2005{\natexlab{a}})}]{bsphiphi}
\bibinfo{author}{\bibfnamefont{D.}~\bibnamefont{Acosta}} \bibnamefont{{\em
  et~al.}} (\bibinfo{collaboration}{CDF Collaboration}),
  \bibinfo{journal}{Phys. Rev. Lett.} \textbf{\bibinfo{volume}{95}},
  \bibinfo{pages}{031801} (\bibinfo{year}{2005}{\natexlab{a}}).

\bibitem[{\citenamefont{Acosta {\em et~al.}}(2005{\natexlab{b}})}]{dcpv}
\bibinfo{author}{\bibfnamefont{D.}~\bibnamefont{Acosta}} \bibnamefont{{\em
  et~al.}} (\bibinfo{collaboration}{CDF Collaboration}),
  \bibinfo{journal}{Phys. Rev. Lett.} \textbf{\bibinfo{volume}{94}},
  \bibinfo{pages}{122001} (\bibinfo{year}{2005}{\natexlab{b}}).

\bibitem[{\citenamefont{Abulencia {\em
  et~al.}}(2006{\natexlab{b}})}]{bcharmless1}
\bibinfo{author}{\bibfnamefont{A.}~\bibnamefont{Abulencia}} \bibnamefont{{\em
  et~al.}} (\bibinfo{collaboration}{CDF Collaboration}),
  \bibinfo{journal}{Phys. Rev. Lett.} \textbf{\bibinfo{volume}{97}},
  \bibinfo{pages}{211802} (\bibinfo{year}{2006}{\natexlab{b}}).

\bibitem[{\citenamefont{Aaltonen {\em et~al.}}(2009)}]{bcharmless2}
\bibinfo{author}{\bibfnamefont{T.}~\bibnamefont{Aaltonen}} \bibnamefont{{\em
  et~al.}} (\bibinfo{collaboration}{CDF Collaboration}),
  \bibinfo{journal}{Phys. Rev. Lett.} \textbf{\bibinfo{volume}{103}},
  \bibinfo{pages}{031801} (\bibinfo{year}{2009}).

\bibitem[{\citenamefont{Aaltonen {\em et~al.}}(2010)}]{mumford}
\bibinfo{author}{\bibfnamefont{T.}~\bibnamefont{Aaltonen}} \bibnamefont{{\em
  et~al.}} (\bibinfo{collaboration}{CDF Collaboration}),
  \bibinfo{journal}{Phys. Rev. Lett.} \textbf{\bibinfo{volume}{104}},
  \bibinfo{pages}{102002} (\bibinfo{year}{2010}).

\bibitem[{\citenamefont{Punzi}(2004)}]{Punzi:2004wh}
\bibinfo{author}{\bibfnamefont{G.}~\bibnamefont{Punzi}} (\bibinfo{year}{2004}),
  \eprint{arXiv:physics/0401045}.

\bibitem[{\citenamefont{Brun {\em et~al.}}(1994)\citenamefont{Brun, Hagelberg,
  Hansroul, and Laselle}}]{GEANT3}
\bibinfo{author}{\bibfnamefont{R.}~\bibnamefont{Brun}},
  \bibinfo{author}{\bibfnamefont{R.}~\bibnamefont{Hagelberg}},
  \bibinfo{author}{\bibfnamefont{M.}~\bibnamefont{Hansroul}}, \bibnamefont{and}
  \bibinfo{author}{\bibfnamefont{J.~C.} \bibnamefont{Laselle}},
  \bibinfo{journal}{Cern programme library, Cern-DD-78-2-REV, Cern-DD-78-2}
  (\bibinfo{year}{1994}).

\bibitem[{\citenamefont{Acosta {\em
  et~al.}}(2005{\natexlab{c}})}]{cdf_detector}
\bibinfo{author}{\bibfnamefont{D.}~\bibnamefont{Acosta}} \bibnamefont{{\em
  et~al.}} (\bibinfo{collaboration}{CDF Collaboration}),
  \bibinfo{journal}{Phys. Rev. D} \textbf{\bibinfo{volume}{71}},
  \bibinfo{pages}{032001} (\bibinfo{year}{2005}{\natexlab{c}}).

\bibitem[{\citenamefont{Affolder {\em et~al.}}(2004)}]{COT}
\bibinfo{author}{\bibfnamefont{T.}~\bibnamefont{Affolder}} \bibnamefont{{\em
  et~al.}}, \bibinfo{journal}{Nucl. Instrum. Methods A}
  \textbf{\bibinfo{volume}{{\bf 526}}}, \bibinfo{pages}{249}
  (\bibinfo{year}{2004}).

\bibitem[{\citenamefont{Sill {\em et~al.}}(2000)}]{SVX}
\bibinfo{author}{\bibfnamefont{A.}~\bibnamefont{Sill}} \bibnamefont{{\em
  et~al.}}, \bibinfo{journal}{Nucl. Instrum. Methods A}
  \textbf{\bibinfo{volume}{{\bf 447}}}, \bibinfo{pages}{1}
  (\bibinfo{year}{2000}).

\bibitem[{\citenamefont{Affolder {\em et~al.}}(2000)}]{ISL}
\bibinfo{author}{\bibfnamefont{A.}~\bibnamefont{Affolder}} \bibnamefont{{\em
  et~al.}}, \bibinfo{journal}{Nucl. Instrum. Methods A}
  \textbf{\bibinfo{volume}{453}}, \bibinfo{pages}{84} (\bibinfo{year}{2000}).

\bibitem[{\citenamefont{Thomson {\em et~al.}}(2002)}]{XFT}
\bibinfo{author}{\bibfnamefont{E.~J.} \bibnamefont{Thomson}} \bibnamefont{{\em
  et~al.}}, \bibinfo{journal}{IEEE Trans. Nucl. Sci.}
  \textbf{\bibinfo{volume}{{\bf 49}}}, \bibinfo{pages}{1063}
  (\bibinfo{year}{2002}).

\bibitem[{\citenamefont{Ashmanskas {\em et~al.}}(2004)}]{SVT}
\bibinfo{author}{\bibfnamefont{W.}~\bibnamefont{Ashmanskas}} \bibnamefont{{\em
  et~al.}}, \bibinfo{journal}{Nucl. Instrum. Methods A}
  \textbf{\bibinfo{volume}{{\bf 518}}}, \bibinfo{pages}{532}
  (\bibinfo{year}{2004}).

\bibitem[{\citenamefont{Amsler {\em et~al.}}(2008)}]{PDG}
\bibinfo{author}{\bibfnamefont{C.}~\bibnamefont{Amsler}} \bibnamefont{{\em
  et~al.}} (\bibinfo{collaboration}{Particle Data Group}),
  \bibinfo{journal}{Phys. Lett. B} \textbf{\bibinfo{volume}{667}},
  \bibinfo{pages}{1} (\bibinfo{year}{2008}).

\bibitem[{\citenamefont{Rademacker}(2007)}]{Rademacker:2005ay}
\bibinfo{author}{\bibfnamefont{J.}~\bibnamefont{Rademacker}},
  \bibinfo{journal}{Nucl. Instrum. Methods A} \textbf{\bibinfo{volume}{570}},
  \bibinfo{pages}{525} (\bibinfo{year}{2007}), \eprint{arXiv:hep-ex/0502042}.

\bibitem[{\citenamefont{Adam {\em et~al.}}(1995)}]{Adam:1995mb}
\bibinfo{author}{\bibfnamefont{W.}~\bibnamefont{Adam}} \bibnamefont{{\em
  et~al.}} (\bibinfo{collaboration}{DELPHI}), \bibinfo{journal}{Z. Phys.}
  \textbf{\bibinfo{volume}{C68}}, \bibinfo{pages}{363} (\bibinfo{year}{1995}).

\bibitem[{\citenamefont{Nason {\em et~al.}}(1988)\citenamefont{Nason, Dawson,
  and Ellis}}]{BGen1}
\bibinfo{author}{\bibfnamefont{P.}~\bibnamefont{Nason}},
  \bibinfo{author}{\bibfnamefont{S.}~\bibnamefont{Dawson}}, \bibnamefont{and}
  \bibinfo{author}{\bibfnamefont{R.~K.} \bibnamefont{Ellis}},
  \bibinfo{journal}{Nucl. Phys.} \textbf{\bibinfo{volume}{B303}},
  \bibinfo{pages}{607} (\bibinfo{year}{1988}).

\bibitem[{\citenamefont{Nason {\em et~al.}}(1989)\citenamefont{Nason, Dawson,
  and Ellis}}]{BGen2}
\bibinfo{author}{\bibfnamefont{P.}~\bibnamefont{Nason}},
  \bibinfo{author}{\bibfnamefont{S.}~\bibnamefont{Dawson}}, \bibnamefont{and}
  \bibinfo{author}{\bibfnamefont{R.~K.} \bibnamefont{Ellis}},
  \bibinfo{journal}{Nucl. Phys.} \textbf{\bibinfo{volume}{B327}},
  \bibinfo{pages}{49} (\bibinfo{year}{1989}).

\bibitem[{\citenamefont{Lange}(2001)}]{EvtGen}
\bibinfo{author}{\bibfnamefont{D.~J.} \bibnamefont{Lange}},
  \bibinfo{journal}{Nucl. Instrum. Methods Phys. Res., Sect. A}
  \textbf{\bibinfo{volume}{462}}, \bibinfo{pages}{152} (\bibinfo{year}{2001}).

\bibitem[{\citenamefont{Fisher}(1936)}]{Fisher1}
\bibinfo{author}{\bibfnamefont{R.~A.} \bibnamefont{Fisher}},
  \bibinfo{journal}{Ann. of Eugenics} \textbf{\bibinfo{volume}{\bf 7}},
  \bibinfo{pages}{179} (\bibinfo{year}{1936}).

\bibitem[{\citenamefont{Abulencia {\em et~al.}}(2007)}]{jpsiLambda}
\bibinfo{author}{\bibfnamefont{A.}~\bibnamefont{Abulencia}} \bibnamefont{{\em
  et~al.}} (\bibinfo{collaboration}{CDF Collaboration}),
  \bibinfo{journal}{\prl} \textbf{\bibinfo{volume}{98}},
  \bibinfo{pages}{122001} (\bibinfo{year}{2007}).

\end{thebibliography}

\end{document}